\documentclass[twocolumn,showpacs,preprintnumbers,amsmath,amssymb]{revtex4}
\usepackage{amsmath}
\usepackage{dcolumn}
\usepackage{bm}
\usepackage{graphicx}

\newcommand{\N}{\mathbb{N}}
\newcommand{\Z}{\mathbb{Z}}
\newcommand{\One}{\openone}
\newcommand{\ke}[1]{| #1 \rangle}

\newcommand{\bk}[2]{\langle #1 | #2 \rangle}
\newcommand{\pr}[2]{|#1\rangle \langle #2|}
\newcommand{\moy}[3]{\langle #1 | #2 | #3 \rangle}
 
\begin{document}
\draft

\title{Tunneling of quantum rotobreathers}

\author{J. Dorignac, S. Flach}

\affiliation{Max-Planck-Institut f\"ur Physik komplexer Systeme,
N\"othnitzer
Stra\ss e 38, D-01187 Dresden}

\date{\today}

\begin{abstract}
We analyze the quantum properties of a system consisting of two nonlinearly
coupled pendula. 
This non-integrable system exhibits two different
symmetries: 
a permutational symmetry (permutation of the pendula) and another one
related to the reversal of the total momentum of the system. 
Each of these symmetries is responsible for the existence of two kinds of
quasi-degenerated states. 
At sufficiently high energy, pairs of symmetry-related states glue together to
form quadruplets. We show that, starting
from the anti-continuous limit, particular
quadruplets allow us to construct quantum states whose properties are very
similar to those of classical rotobreathers.
By diagonalizing numerically the quantum Hamiltonian, we investigate their
properties and show that
such states are able to store the main part of the total energy on one of
the pendula. Contrary to the classical situation, the coupling 
between pendula necessarily introduces a periodic exchange of energy between
them with a frequency which is proportional to 
the energy splitting between quasi-degenerated states related to the
permutation symmetry. This splitting may remain very small
as the coupling strength increases and is a decreasing function of 
the pair energy. The energy may be
therefore stored in one pendulum during a time period very long as compared to
the inverse of the internal rotobreather frequency.     
 
\end{abstract}

\maketitle

\vskip2pc
\narrowtext

\section{Introduction}
 
As revealed by the increasing number of recent papers
\cite{Ovchinnikov70}-\cite{Voulgarakis01}, the quantum counterpart or the 
quantization of classical discrete 
breathers (CDBs) has become these very last years a real 
challenging and exciting
field. At present, indeed, the theory of CDBs, i.e. "time-periodic spatially
localized motions in networks of oscillators (MacKay \cite{MacKay00})", 
has reached a high degree of perfection and may certainly be considered as a
real achievement. As it is well known, the necessary condition required
for a Hamiltonian lattice to support such CDBs is that the breather frequency
as well as its harmonics do not lie inside the phonon band. Such a non
resonance condition may be achieved by the interplay of discreteness and
nonlinearity, the first providing a natural upper bound of
the linearized Hamiltonian spectrum (say phonons) whereas the latter 
allowing for a tuning of frequencies
out of this spectrum \cite{Flach94}. 

Mathematical existence proofs 
of CDBs have been obtained for a wide variety of model Hamiltonians
\cite{MacKay94} and have simultaneously given rise to an accurate
numerical method allowing for their practical construction \cite{Marin96}. 
Both are based on the
so-called {\em anti-continuous} limit (i.e. the limit where the coupling
between oscillators vanishes),
provided a nonlinear onsite potential
is present. In this limit
time-periodic spatially localized (on one
or more sites) solutions trivially exist. 
More recently, an existence proof of
CDBs based on a discrete version of the center manifold reduction has also
been given for the FPU chain for which the method based on 
the anti-continuous limit fails \cite{James01}. 
This completes and enlarges the results previously obtained in
\cite{Flach95} by a homoclinic orbit approach. At last, also recently, a 
variational approach has been carried out to prove the existence of hard
discrete breathers in some classes of Hamiltonians \cite{Aubry01}. 
Obviously, the wide range of applications of these rigorous
mathematical results supports the idea that CDBs are generic solutions
of nonlinear Hamiltonian lattices as claimed for the first time 
by Sievers and Takeno in 1988 when they discovered this new kind of intrinsic
localized modes \cite{Sievers88}. 

Together with their existence proofs, the properties of CDBs have been
extensively studied. It has been shown for example 
that they are structurally stable
provided the non resonance condition holds \cite{Flach94} and linearly stable 
in any dimension provided the coupling is weak enough \cite{Aubry97}. Their
spatial decay is generally exponential (for any finite range interaction
potential) but may be also algebraic in case the interaction potential itself
is algebraically decaying \cite{FlachPRE98}. More recently, it has even been
confirmed that a pure nonlinearity of the interaction 
potential may give rise to a
super-exponential decay of the breather tail
\cite{Flach94,Dey01}. 
At last, and directly related to their spatial decay
properties, possible energy thresholds for their appearance have been 
derived according to whether the lattice dimension exceeds  
a system-dependent critical value or not 
\cite{Flach97}. Extensive studies of CDBs properties as well as the
perspectives in this field may be found in different reviews 
(see for instance \cite{Aubry97,Flachrep98,MacKay00,AubryNATO01,FlachZol01}).
       
The general properties of CDBs briefly listed above shows how well 
their theory is
understood at present. Much less is known on the other hand about their
quantum counterparts (QDBs), that is, about
the quantum states which would behave
similarly to CDBs in a sense to be clarified below. 
Nevertheless, this question is of great importance given
one natural area of application, namely, the condensed matter physics
where quantum effects have generally to be considered (see \cite{MacKay00} and
refs. therein for a review of the
materials where the presence of DBs is suspected, expected or 
has been detected). This is the precise
reason why the quantum theory of breathers is currently in full development.
       
Let us turn back to an essential property of CDBs. These time-periodic
solutions occur in a lattice but are spatially localized which means that
they are not invariant under the discrete translational symmetry of the
lattice whereas the Hamiltonian does. If we now consider the 
corresponding quantum problem, 
the invariance of the quantum Hamiltonian $\hat{H}$
with respect to the discrete translational symmetry (the operator of which we
denote by $\hat{T}$) yields $[\hat{H},\hat{T}]=0$. The eigenstates of
$\hat{T}$ are thus eigenstates of $\hat{H}$ and are delocalized 
along the lattice (Bloch waves). Of course, this discrete
translational symmetry of the lattice is broken if we consider for instance a
finite system with fixed or open boundary conditions. 
Nevertheless, if the lattice consists of a
sufficiently large number of sites, discrete translational symmetry
is practically restored similar to the  
infinite lattice. However, 
at large enough energy the quantum and the classical descriptions of the system
should give similar results!  

It is possible to reconcile the quantum and the classical points 
of view provided 
we again start from the anti-continuous limit. In this limit indeed, the
Hamiltonian consists only of a sum of identical Hamiltonians (one for each
site) $\hat{H} = \sum_s \hat{H}_s$ and a general eigenstate $\ke{\psi}$ of the
system may be represented as the tensorial product of local eigenstates
$\ke{\phi_s}$, $\ke{\psi}= \otimes_s \ke{\phi_s}$. Then, a
localized excitation is obviously constructed by creating
an excitation of level $k$ at site $n$, $\ke{\phi^{(k)}_n}$ whereas
the other sites reside in their ground state $\ke{\phi^{(0)}_{s \neq n}}$.
This leads to $\ke{\psi^{(k)}_n} = \otimes_{s<n} \ke{\phi^{(0)}_s}
\otimes \ke{\phi^{(k)}_n} \otimes_{s>n} \ke{\phi^{(0)}_s}$.
This case corrresponds to the classical one used to initialize the construction
algorithm of a CDB at non zero coupling. Nevertheless,
any eigenstate of the form $\ke{\psi^{(k)}_j} = \otimes_{s<j}
\ke{\phi^{(0)}_s} \otimes \ke{\phi^{(k)}_j} \otimes_{s>j} \ke{\phi^{(0)}_s}$
has the same energy. The
corresponding eigen-subspace has a dimension $N$ equal to the number of sites
(provided no additional accidental degeneracy occurs). Therefore in the
uncoupled (anti-continuous) limit, any unitary transform of the preceeding
basis leaves the subspace invariant although it yields a new basis. As the
Hamiltonian system is invariant under the discrete translation
$\hat{T}$ whatever the intersite coupling $\varepsilon$ is, at non
zero coupling, the eigenstates of $\hat{H}$ must belong to one of the $N$
symmetry sectors defined by $\hat{T}$. When the coupling becomes zero, we may
thus choose the eigenbasis of $\hat{T}$ as a basis of the $N$-dimensional
subspace. We denote its eigenvectors by
$\ke{\lambda^{(k)}_q}$ where $q$ labels the symmetry sector. 
Of course each of these new eigenstates is completely delocalized (in the
sense that $\forall q,\ [\sum_j
|\bk{\psi^{(k)}_j}{\lambda^{(k)}_q}|^4]^{-1}=N$).  

Let us assume now that we switch on the coupling between the sites. We expect
the $N$-fold degenerated eigenenergy to be splitted under the effect of the
perturbation and to form a band of $N$ nearly degenerated eigenenergies.  
The corresponding eigenstates, which already had the correct symmetry
because of the
unitary transform $\hat{T}$ performed on the local basis at $\varepsilon = 0$,
now give 
a new (perturbed) basis $\ke{\lambda^{(k)}_q(\varepsilon)}$. 
Provided $\varepsilon$ is weak enough, these
new eigenvectors are close to those defined at the uncoupled limit and thus an
inverse unitary transform $\hat{T}^{-1}$ is expected to yield a basis 
$\ke{\psi^{(k)}_n(\varepsilon)}$ close to the local basis of the uncoupled
limit, that is, $\ke{\psi^{(k)}_n(\varepsilon)}= \ke{\psi^{(k)}_n} +{\cal
O}(\varepsilon)$. These states are thus well localized provided the coupling
is weak enough. However, being a linear combination of non degenerated
eigenstates, they are no more eigenstates of $\hat{H}$ and evolve in
time. This tunneling effect, corresponding to the transfer of an initial
excitation from site to site, takes a time typically given by the width of the
band at a given value of the coupling $\varepsilon$. In order to gap the
bridge between the classical manifestation of the breather solution and its
quantum realization, we thus expect the quantum system to exhibit bands of
$N$ nearly degenerated eigenstates whose width tends to zero as the average
energy of the band goes to infinity.

The purpose of a quantum theory of breathers is thus to know how the above 
defined bands behave as the coupling increases. Several successful attempts
to answer this question have already been done in one-dimensional systems
which present the particularity to be integrable
(\cite{Bernstein90}, \cite{Salerno92}-\cite{Scott94} and the review
\cite{Scott00}). The 
(soliton/breather) bandwidth is shown to behave typically like $\Delta E_n \sim
(\varepsilon/\gamma)^n/(n-1)!$  where $n$ is 
strictly speaking the number of bosons in the system. This number
corresponds to the number of the excited level of a single site
in the absence of interaction $\varepsilon=0$ and typically depends
on energy as a power law.
Finally $\gamma$ is a measure of the onsite (quartic)
nonlinearity.  It is then obvious that for bands of high energy ($n
\rightarrow \infty$), the bandwidth goes rapidly
(in fact more than exponentially fast) to zero
with increasing $n$. In this case the tunneling time goes to infinity in the classical
limit and the quantum breather (quasi-classical) state remains localized on
its initial site. 

Several questions remain however. At high energy, if the
onsite potential behaves like $x^{2q}$, the density of states (evaluated at
$\epsilon = 0$ by the mean of the Weyl's formula) scales like
$g(E) \sim E^{N(1+q)/(2q)-1}$ as $E \rightarrow \infty$, where $N$ is the
number of sites. An immediate conclusion is that if the number of sites is
greater or equal to 2, the density of states (DOS) 
increases with energy according to
a power law with exponents linearly depending on $N$.
Then, we may ask whether the corresponding 
raise of interaction possibilities (hybridization) will destroy the
quantum breather bands or not. A partial answer to this question should be given by
the ratio of the breather bandwidth and the mean level spacing $1/g(E)$ at a
given (high) energy $E$: $\lim_{E \rightarrow \infty}g(E)\Delta E$ ? 
Assuming that the DOS scales as indicated by the Weyl's formula,
the ratio of the bandwidth to the mean level spacing will tend
to zero provided the bandwidth decreases exponentially with
increasing energy (or faster).
The above mentioned integrable examples fall into this category.
This result seems to indicate and to explain the
possibility, starting from the quantum problem, to recover the limit of
the classical breather at high energy. Another natural question is related to 
the impact of the nonintegrability of the system on the quantum 
breather bands and in
particular the influence at the quantum level of the chaotic
trajectories induced by the nonintegrability lying nearby the classical
breathers. 

Up to now, except for integrable systems, where analytical results are
obtainable, studies of quantum breathers have been done numerically.
Such studies become rapidly extremely difficult due to the
huge matrices to be diagonalized and so far have been restricted to small 1D
systems for which the number of sites has not exceeded $N=12$
\cite{Kladko99}. Even for these moderate lattice sizes, the average dimension
of the (truncated) Hamiltonian matrices is 
generally of order $10^6$ and requires specific
numerical diagonalization methods \cite{Wang96,Wang98}. 
The numerical results reported in the papers mentioned above are
restricted to the low energy sector as soon as the number of sites exceeds a
few units. This is easily understood given the number of configurations
obtainable by truncating the basis to $p$ bosons per sites on $N$ sites
($(p+1)^N$). Remains the possibility to study very small systems ($N=2$ or
$3$), as it has already been done for the dimer (integrable)
\cite{Bernstein90,Aubry96} and the trimer (non integrable)
\cite{FlachFle97,FlachFle01}. The dimer has also been used to
describe the tunneling of a QB along a chain by a suitable linearization of
the lattice around it \cite{Fleurov98}. This linearization method was
also employed to study the properties of classical rotobreathers in a 
chain of pendula \cite{Takeno97}. 
But the specific properties of quantum rotobreathers have not been
discussed yet.
We will consider a dimer of two coupled quantum pendula which, 
due to the presence of both nonlinear onsite and
interaction potentials, is not integrable.  

As far as we know, no experimental study of quantum rotobreathers has
been done yet. This is possibly due to their quite large activation energy.
The reader should notice however that a large number of studies have been
devoted to the rotational motions of molecules (see
e.g. \cite{Ovchinnikov01,Svensson99,Brown00,Havighorst96})
especially concerning the methyl groups whose dynamical properties are usually
obtained via neutron scattering
\cite{Fillaux90,Fillaux91,Fillaux98,Colmenero98,Colmenero00}. 
But the transitions thus obtained concern the so-called quantum rotational
tunneling effect \cite{Colmenero00}. This tunneling occurs between the
equilibrium positions defined by the onsite potential which is $n$-fold
according to the individual symmetry of the observed molecule and the symmetry
of its environment. For the methyl groups $n$ is generally equal to 3 but for
ammonia in Hofmann clathrates, the fourfold symmetry of the host crystal
induces an approximative 12-fold \cite{Wegener90} or more complex
\cite{Havighorst96,Kolarschik96} symmetry. This tunneling is thus responsible
for a rotation of the molecule forbidden in the classical case. The quantum
rotobreather on an other hand is a state whose energy is situated {\em above}
the energy of the separatrix. A tunneling effect appears, because of the
coupling between molecules, which consists in 
the transfer of the excitation from site (molecule) to site. 
A study of the properties of the 4-methyl-pyridine by Fillaux and co-workers
has revealed the presence of a quantum sine-Gordon breather in this
compound \cite{Fillaux90,Fillaux91,Fillaux98}. 
The properties of this state, shown to be the ground state of the
system, are theoretically analyzed via a semiclassical quantization procedure
of the classical solution of the sine-Gordon equation and then successfully 
compared to experimental results. However, the coupling between adjacent
methyl groups is so strong along certain axes of the crystal (chains) 
that the relative phases on neighbouring groups are small. 
That allows to use the continuum sine-Gordon theory.
In the
case of a rotobreather, such an approximation becomes of course impossible due
to the unavoidable large phase difference created at the interface between 
the rotating and oscillating groups. The study of the properties of 
rotobreathers (either classical or quantum) thus requires to preserve the
rotational invariance of the interaction potential.

Our system under study consists
of two pendula coupled by a cosine interaction potential (which preserves
the rotational invariance). To simplify the study, the onsite and interaction
potentials are the same functions and are 1-fold. This avoids to deal with the
nearly degeneracies coming from the $n$ different arches of a $cos(nx)$
potential at low energies (typically below the separatrix). Notice that this
does not change essentially the high energy 
part of the single site spectrum (which
tends to the free rotor one whatever the value of $n$) 
but removes the splittings of the doublets which
are not coming from the center ($k=0$) or the edge ($k=n/2$) of the
corresponding Brillouin zone $k \in \{-n/2+1,..,n/2\}$. The results we will
obtain in section 2 concerning the upper part (above the separatrix energy) of
the spectrum are thus easy to adapt to the case
of an $n$-fold onsite potential. Notice at
last that despite its very short size, our "pendula dimer" still possesses
the discrete translational invariance $\hat{T}$ which corresponds 
to the permutation symmetry.

The plan of the paper is as follows: in section 2 we briefly review the
properties of a single quantum pendulum and we derive important formulas
concerning the splitting occurring in the doublets in the high energy
sector. In
section 3, we start by presenting some results of the classical problem
corresponding to two coupled pendula. Then we show how to compute the
quantum spectrum of this system. We derive the exact spectrum at the uncoupled
limit and we construct a quantum rotobreather according to the method
presented in the introduction. By increasing the coupling between the
pendula, we follow the quantum rotobreather (QR) 
and we compute the splittings
occurring in the corresponding quadruplet. 

  
\section{Single pendulum problem (SPP)}

In this section we briefly review some of the properties of the classical and
quantum one pendulum problem (for more details see for instance
\cite{Aldro80}). In the quantum case, we particularly focus on
quasi-degenerated states and their energy splitting. We show by two different
methods that it is possible to compute this splitting exactly in leading
order. The Hamiltonian of the pendulum system is given by  
\begin{equation} \label{HP1}
{\cal H} (x,p) = \frac{1}{2} p^2 + \alpha (1-\cos (x))  
\end{equation}
where $x$ and $p$ represent respectively the angle variable and the associated
momentum. $\alpha > 0$ tunes the barrier height of the onsite potential.  
\subsection{Classical case}
From (\ref{HP1}) we deduce the classical equation of motion
\begin{equation} \label{EM1}
\ddot{x} + \alpha \sin (x) = 0  
\end{equation}
which possesses two different solutions. Denoting by $E$ the
energy of the pendulum 
and by $E_s = 2\alpha $ the energy of the separatrix separating the oscillatory
motion from the rotational one we get:

\begin{itemize}
\item Oscillation: $E < E_s$
\begin{equation} 
\sin (x/2) = k {\rm sn}\left(\sqrt{\alpha} t ; k \right)
\end{equation}
where the modulus of the Jacobian elliptic function ${\rm sn}$ is defined by
the relation $k = (E/2\alpha)^{1/2}$ and the
period of the oscillation is $T_{\rm osc} = 4 K(k)/\sqrt{\alpha}$, $K(k)$ 
being the complete elliptic integral of the first kind \cite{Abr};

\item Rotation: $E > E_s$
\begin{equation} 
\sin (x/2) = {\rm sn}\left(\frac{\sqrt{\alpha}}{\tilde{k}} t ; \tilde{k}
\right)       
\end{equation}
where $ \tilde{k} = (2\alpha/E)^{1/2} $ and where the rotation
period is $T_{\rm rot} = 2 \tilde{k} K(\tilde{k})/\sqrt{\alpha}$. 

\end{itemize}

\subsection{Quantum case}

Surprisingly, while the classical problem of the pendulum is a very basic one, 
it seems that a thorough study of the related quantum
problem has been done quite recently
(R. Aldrovandi and P. Leal Ferreira \cite{Aldro80}). Our purpose here is not
to repeat the analysis done in the paper cited above but rather to insist on
the momentum reversal symmetry of the Hamiltonian (\ref{HP1}) which leads to
the appearance of pairs of quasi-degenerated states above the
separatrix.

\subsubsection{Analytical solution}

The stationary Schr\"odinger equation corresponding to the Hamiltonian
(\ref{HP1}) is given by
\begin{equation} \label{Schr1} 
-\frac{1}{2}\frac{d^2 \psi(x)}{d x^2}+\alpha (1-\cos(x)) \psi(x) = E \psi(x).
\end{equation}   
$\psi$ is the wave function of the pendulum and $E$ its energy. As the wave
function of the pendulum has to be single-valued
(\cite{Aldro80},\cite{Ross00}) , we impose the periodicity condition 
\begin{equation} \label{Percond1}
\psi(x+2\pi) = \psi(x).
\end{equation}
It is possible to get an analytical solution of (\ref{Schr1}) by performing the
following change of variables: $u=(\pi-x)/2$ and $\phi(u)=\psi(x)$.
We immediately obtain the canonical form of the Mathieu equation (\cite{Abr})
\begin{equation} \label{Mathieu} 
\frac{d^2 \phi(u)}{d u^2}+(a -2 q \cos(2u)) \phi(u) = 0
\end{equation}
where
\begin{equation} \label{defqa}
q = 4 \alpha \ \ \ \ {\rm and}\ \ \ \ a = 8 (E-\alpha).
\end{equation}
Because of the previous change of variables, $\phi(u)$ is now a $\pi$-periodic
function. It can be shown (\cite{Abr}) that the Mathieu equation supports
$\pi$-periodic solutions if and only if the characteristic value $a$ belongs
to an infinite countable set of values denoted by $\{a_{2n}(q),b_{2n}(q)\}\ \
n \in \N $. $a_{2n}$ is related to the {\em even} Mathieu function
$ce_{2n}(u,q)$ 
whereas  $b_{2n}$ is related to the {\em odd} Mathieu function $se_{2n}(u,q)$.
As follows
from their definition (\ref{defqa}), $q$ is directly related to the energy of
the separatrix $E_s = 2\alpha$ that is to the depth of the cosine potential
appearing in the Schr\"odinger equation whereas, up to a scaling and a shift
factor,  $a$ represents the eigenenergies of the pendulum Hamiltonian. For a
given value of $q$ (that is of $\alpha$), the series $\{a_{2n},b_{2n}\}$
increases monotonically with $n$ keeping the property 
$b_{2n} < a_{2n}$. For $n = 0$ the
only possible solution is an even function $ce_{0}(u,q)$ which represents the
ground state of the system. 

According to the previous results, the analytical solution of the
Schr\"odinger equation (\ref{Schr1}) reads
\begin{eqnarray} \label{Psimath}
  \psi_{2n}^{(e)} =
 \frac{1}{\sqrt{\pi}}ce_{2n}\left(\frac{\pi-x}{2},q\right) \ &;&\
 E_{2n}^{(e)} = \frac{a_{2n}(q)}{8} + \alpha \nonumber \\
 \psi_{2n}^{(o)} =
 \frac{1}{\sqrt{\pi}}se_{2n}\left(\frac{\pi-x}{2},q\right) \ &;&\
 E_{2n}^{(o)} = \frac{b_{2n}(q)}{8} + \alpha 
\end{eqnarray}
for $n \in \N^*$ and
\begin{equation}
\psi_{0}^{(e)} =
 \frac{1}{\sqrt{\pi}}ce_{0}\left(\frac{\pi-x}{2},q\right) \ \ \ ;\ \ \
 E_{0}^{(e)} = \frac{a_{0}(q)}{8} + \alpha 
\end{equation}
for $n = 0$ which represents the ground state of the pendulum Hamiltonian.

\subsubsection{The Fourier representation (FR)}

According to the periodicity condition (\ref{Percond1}) imposed to the wave
function, it is rather natural to consider its Fourier expansion: 
\begin{equation} \label{devFour}
\psi(x) = \frac{1}{\sqrt{2\pi}}\sum_{m \in \Z} \psi_m e^{i m x}
\end{equation}
where 
\begin{equation}
\psi_m = \frac{1}{\sqrt{2\pi}}\int_0^{2\pi} \psi(x) e^{-i m x} dx .
\end{equation}
Because of the periodicity of the wave function, the Fourier space associated
to the pendulum problem is infinite but discrete. In the FR, the
stationary Schr\"odinger equation becomes
\begin{equation} \label{Schfour}
 m^2 \psi_m -\alpha (\psi_{m+1} + \psi_{m-1}) = \tilde{E}\psi_m \ \ \ , \ 
\ \ m \in \Z
\end{equation}
where for later convenience we have redefined the energy as
\begin{equation}
\tilde{E} = 2 (E-\alpha) .
\end{equation}

Equation (\ref{Schfour}) represents a tight-binding equation whose
hopping terms and onsite potential would be respectively $-\alpha$ and $m^2$.
In this equation, the sum $(\psi_{m+1} + \psi_{m-1})$ related to the cosine
potential plays
now the role of a discrete Laplacian while the onsite "potential" term $m^2$
comes from the kinetic energy. Thus, in FR,
the terms of the
Schr\"odinger equation have inverted their role. This fact is important in
understanding how 
the Discrete
Wentzel-Kramers-Brillouin (DWKB) method  (\cite{Braun93},\cite{Garg98})
applies to the Mathieu equation (see section \ref{sectDWKB}). 

Equation (\ref{Schfour}) is invariant under the transformation ($m \rightarrow
-m$).  This symmetry operation corresponds to the inversion of the momentum of
the system ($p \rightarrow -p$) which originates in
the time reversal symmetry of the original problem.
This allows us to separate the eigenstates of
(\ref{Schfour}) into symmetric $\ke{s}$ and antisymmetric $\ke{a}$ states.
  
We may write the Hamiltonian of (\ref{Schfour}) as
\begin{equation} \label{Hfour1}
  \tilde{H} = \sum_{m \in \Z} m^2 \pr{m}{m} -\alpha (\pr{m+1}{m} + \pr{m}{m+1})
\end{equation}     
where we have used the ket notation $\ke{m}$ to represent the plane wave
function $\bk{x}{m}=e^{imx}/\sqrt{2\pi}$. 
The matrix representing $\tilde{H}$ in the FR is infinite, tridiagonal and
symmetric. Its diagonal elements
are $m^2$ and the
off-diagonal elements are constant and equal to $-\alpha$. Moreover,
$\tilde{H}_{m,n}=\tilde{H}_{-m,-n}$. This additional "central" symmetry is a
direct consequence of the time reversal symmetry and the Hermitian
properties of $\tilde{H}$.

\subsubsection{Low-energy states}

Without any onsite
potential ($\alpha = 0$), it follows from (\ref{Hfour1}) that $\ke{m}$ and
$\ke{-m}$ are eigenstates of $\tilde{H}$ with identical energy $m^2$. 
Consequently, each state except the ground state $\ke{0}$ is
two-fold degenerated in this limit. 
This two-fold degeneracy is due to the two equivalent
possible motions consisting in rotating in a given sense or its opposite.

\begin{figure}[htbp]
\includegraphics[width=0.48\textwidth]{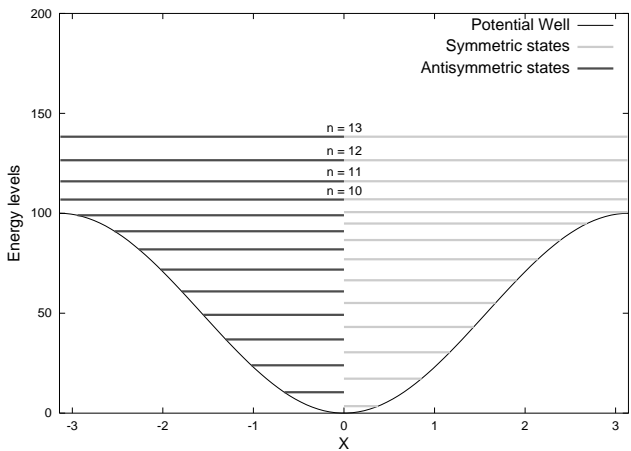}
   \caption{Energy levels of a pendulum with $\alpha=50$. 
States have been distinguished according to their parity. Only
   a few states lying above the separatrix have been displayed. The number $n$
   of the corresponding doublet is written on top of it.
   \label{ELalp5e1}} 
\end{figure}

The switching on of an onsite potential governed by the parameter $\alpha$
lifts this degeneracy but in a way depending
on the energy level of the state
under consideration. Indeed, when $\alpha$ becomes nonzero, the pendulum
system admits a new kind of motion, namely, the oscillating
motion which corresponds to the motion
in the cosine arch of the potential well in the classical system 
(see Fig. \ref{ELalp5e1}). 
The quantum system also admits such
kind of states but their number is limited
by the value of $\alpha$ because of the quantization rules. 
As $\alpha$ becomes larger, the number of states
below the separatrix increases and can be estimated to $8\sqrt{\alpha}/\pi$
by using the Weyl formula (see below (\ref{nEOPP}) in the limit $k \rightarrow
1$). If the value of
$\alpha$ is large enough it becomes possible to look at the properties of the
low energy states by expanding the cosine potential around zero in a series of
powers of $x$. We obtain
\begin{equation} \label{cosexpan}
\alpha (1-\cos (x)) = \alpha \left( \frac{1}{2} x^2 - \frac{1}{24} x^4
+\frac{1}{720} x^6 +\circ (x^7)\right) .
\end{equation}

The first order of this expansion leads to the harmonic
approximation of (\ref{HP1}) whose frequency is given by $\omega =
\sqrt{\alpha}$. Thus, low energy states of the quantum pendulum are well
represented by the corresponding low energy harmonic eigenstates (at
least for large enough values of $\alpha$). As the level spacing of the 
harmonic oscillator is constant (and equal to
$\omega$) we expect the states to be quite regularly spaced deep inside
the well. This is illustrated in Fig. \ref{ELalp5e1} which represents
the lower part of the spectrum of a pendulum
whose parameter $\alpha =50$. The energy levels of $\ke{s}$ and $\ke{a}$
states have been respectively represented to the right and to the left.
 We clearly see that the few first states 
are quite regularly spaced while 
they condensate in reaching energies close to the separatrix. 
By using standard second order perturbation theory
within the natural harmonic basis, 
we may obtain the first corrections to the pure harmonic spectrum. This
perturbation in the nonlinear terms gives
\begin{eqnarray} \label{Lowen}
E_n &=& \frac{1}{2}(n+(n+1))\omega - \frac{1}{2^5}(n^2+(n+1)^2) \nonumber \\
 &-& \frac{1}{2^9}(n^3+(n+1)^3)\frac{1}{\omega}
+{\cal O}\left(\frac{1}{\omega^2}\right) 
\end{eqnarray}
where $n$ labels the states ($n \in \N$).
Of course, this expression ceases to be valid when nonlinear corrections
become large as compared to the harmonic term. Nevertheless, it shows how 
the spacing becomes smaller as the energy increases but remains below 
the energy of the separatrix:
\begin{eqnarray} 
\Delta_n &=&  E_{n+1}-E_n \simeq \left(\omega -
\frac{1}{8}-\frac{1}{64\omega}\right ) \nonumber \\
&-& \left(\frac{1}{8}+\frac{3}{128\omega}\right)n-\frac{3}{256\omega}n^2 .
\end{eqnarray}

\subsubsection{High-energy states}

Above the separatrix, Fig. \ref{ELalp5e1} shows that symmetric and
antisymmetric states glue together to form pairs of nearly degenerated states.
From a physical point of view this phenomenon is due to the fact that far
above from the separatrix the cosine potential appears like a perturbation of
the free rotor and only shifts the levels a bit 
around their value $\tilde{E}^{\rm
free \, rotor}_m=m^2$ 
($m^2 \gg 2\alpha$). Because of the momentum reversal symmetry, the Hamiltonian
(\ref{Hfour1}) diagonalizes into two blocks each related respectively to
$\ke{s}$ or $\ke{a}$ states. In the FR, the corresponding reduced
matrices are still tridiagonal. By applying standard perturbation theory in
the parameter $\alpha$ to one of these matrices whose spectrum is now free
of nearly degenerated eigenvalues, we obtain corrections to
the free rotor energy $\tilde{E}^{\rm free \,rotor}_m=m^2$. To second order,
we find
\begin{equation} \label{Endoub}
\tilde{E}^d_n = n^2 + \frac{2 \alpha^2}{4n^2-1} + {\cal
O}\left(\frac{\alpha^4}{n^6}\right) .
\end{equation}
Explicit corrections up to $\alpha^6/n^{10}$ are obtainable (see \cite{Abr}
p. 724). As we mentioned, this energy is the mean energy of a doublet. This
is indicated by the subscript ${}^d$ in (\ref{Endoub}).

\subsubsection{Analytical computation of the splitting: High-order
perturbation theory}

It is also possible to calculate explicitely the splitting between the
symmetric and antisymmetric states of a doublet labeled by $m$ ($m^2 \gg
2\alpha$). The inverse of this splitting
is a direct measure of the time required by a rotating state to invert its
initial momentum. It has to be noticed that by "rotating state" we mean the
superposition of the states $\ke{s}$ and $\ke{a}$ belonging to the same
doublet. Because of their symmetry, the expectation value of the momentum for 
$\ke{s}$ and $\ke{a}$ are zero. 

A possible way to get this splitting is to use high order perturbation theory
as it has already been done in \cite{Aubry96} (equation (12)). A proof of this
formula has been derived in \cite{Bernstein90} and applied to the quantum
discrete self-trapping equation. 
The special "centro-symmetric" form of the tridiagonal Hamiltonian matrix 
in the FR allows us to compute the exact leading order of the splitting as
  
\begin{equation} \label{DEn}
{\Delta E}_n =  \frac{1}{2} {\Delta \tilde{E}}_n =
\frac{1}{8}(a_{2n}(4\alpha)-b_{2n}(4\alpha)) = \frac{\alpha^{2n}}{(2n-1)!^2} .
\end{equation}
Notice that due to the obvious relation between the eigenfunctions of the SPP
and the Mathieu functions, the splitting ${\Delta E}_n$ 
gives in turn the result of the splitting between the
characteristic values associated to the symmetric and antisymmetric
$\pi$-periodic Mathieu functions $ce_{2n}$ and $se_{2n}$. 
A similar computation allows us to obtain the splitting for the $2\pi$-periodic
Mathieu functions. By using the notations of \cite{Abr}, we get
\begin{equation} \label{ambMat}
a_{r}(q)-b_{r}(q) =  \frac{8}{(r-1)!^2}\left(\frac{q}{4}\right)^r + o
\left(q^{r+1}\right)  \ \ \ \text{as}\ \ \ r \rightarrow \infty .
\end{equation}  
To our knowledge this is the first time that such a splitting
has been exactly computed in leading order \footnote{
It is also possible to compute the first correction to the splitting itself
either by using the Rayleigh-Schr\"odinger perturbation theory applied to each
symmetry sectors \cite{DoriFlach} or by using the Brillouin-Wigner method
\cite{Ulyanov99}. We have obtained
$$ a_{r}(q)-b_{r}(q) =
\frac{8}{(r-1)!^2}\left(\frac{q}{4}\right)^r \left[ 1 -
4\left(\frac{q}{4}\right)^2\frac{r+2}{(r^2-1)^2} \right] . $$
For $r=1$, the correction to the leading order has to be computed separately. 
It reads $-q^3/32$. This expression is in exact agreement with the small $q$
expansion of the characteristic values. }. 
In particular, it corrects
and precises the formula obtained in \cite{Abr}. Moreover, it can be verified
to be correct by using the small $q$ expansions of the low order 
characteristic values themselves.  

As $n$, which represents the label of the doublet, becomes large enough,
we find
the following asymptotic form for the splitting
\begin{equation} \label{AsDEn}
{\Delta E}_n \sim
\left(\frac{n}{\pi}-\frac{1}{12\pi}\right)
\left(\frac{e\sqrt{\alpha}}{2n}\right)^{4n} \ \ \ (n \rightarrow \infty)  
\end{equation}  
indicating that it decays more than exponentially fast as $n$
increases. Moreover this expression contains its limit of validity as
the number $n$ has to be greater than the critical value $n_c =
e\sqrt{\alpha}/2$ to ensure that the splitting is small (see comment 
\footnote{It is
interesting to compare this result with the one we may obtain by evaluating
the number of states below the separatrix $N_s \simeq 8\sqrt{\alpha}/\pi$,
$\alpha \gg 1$  (see (\ref{nEOPP}) in the limit $k \rightarrow 1$).
The number of (nondegenerated) pairs of symmetric and antisymmetric
eigenstates below the separatrix is thus $n'_c \simeq N_s/2 =
4\sqrt{\alpha}/\pi$ which is very similar to $n_c$ as $8e/\pi \simeq
1.067$. As $E_{N_s} \simeq E_s = 2\alpha$ it proves that expression
(\ref{DEn}) starts to be valid immediately as $E > E_s$.}).  
 
\subsubsection{Discrete WKB theory} \label{sectDWKB}

\begin{figure}[htbp]
\includegraphics[width=0.48\textwidth]{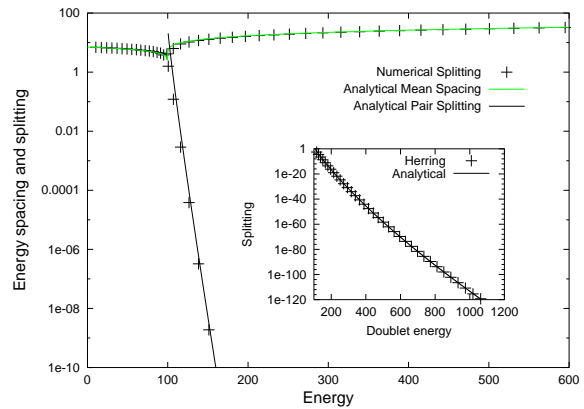}
   \caption{Spacings and splittings of the spectrum of a quantum pendulum for 
   $\alpha=50$. Quantities are described in the text.
   \label{SPAcSPLi}} 
\end{figure}

Another way to compute the doublet splitting is to use the DWKB method
developed in \cite{Braun93} and to use the discrete counterpart
of the Herring's
formula as done in \cite{Garg98}. Following \cite{Garg98} and according to
(\ref{Schfour}), the splitting between the
nearly degenerated states reads
\begin{equation} \label{Herring}
\Delta E_n = \frac{1}{2} |\tilde{E}_s^{(n)}-\tilde{E}_a^{(n)}| = \alpha
s^{(n)}_0 a^{(n)}_1  
\end{equation} 
where $s^{(n)}_0$ and $a^{(n)}_1$ are the Fourier components 
of the symmetric and antisymmetric states of the $n$-th doublet. 
Instead of evaluating 
these components by using the connection formulae of \cite{Braun93},
we use directly the resulting forms given 
in \cite{Abr}. The asymptotic behavior (see expression {\bf 20.2.29} 
of \cite{Abr}) of the
Fourier components of the Mathieu functions yields
\begin{eqnarray}
A_0^{(2n)}(q) &=& \frac{(n-1)!}{n!(2n-1)!} \frac{q^n}{4^n} A_{2n}^{2n}(q) 
\nonumber \\
B_2^{(2n)}(q) &=& \frac{n!}{(n-1)!(2n-1)!} \frac{q^{n-1}}{4^{n-1}}
B_{2n}^{2n}(q) \nonumber  
\end{eqnarray} 
where $A$ and $B$ are the coefficients of the cosine and sine Fourier series of
the $\pi$-periodic Mathieu functions $ce_{2n}$ and $se_{2n}$. The superscript
indicates the order whereas the subscript 
labels the Fourier components. The correct 
normalization 
yields $s^{(n)}_0 = \sqrt{2} A_0^{(2n)}(q)$ and $a^{(n)}_1 =
B_2^{(2n)}(q)/\sqrt{2}$ (see (\ref{Psimath}) and
 (\ref{devFour})). By taking into account that $A_{2n}^{2n}(q) =
B_{2n}^{2n}(q) = 1 + o(q)$ and $\alpha=q/4$, we finally get
\begin{equation} \label{DeltanDWKB}
\Delta E_n = \frac{\alpha^{2n}}{[(2n-1)!]^2}
\end{equation}
which coincides with (\ref{DEn}). The graph (Fig. \ref{SPAcSPLi})
shows the 
spacings and the splittings of the SPP spectrum for $\alpha=50$ as a function
of the energy $E$. The
crosses correspond to $E_{n+1}-E_{n}$ as obtained
from the numerical diagonalization of 
(\ref{Schfour}). The two upper
branches (dotted line) represent the spacing between {\em neighboring
eigenvalues} (when $E<E_s$) and the spacing between the {\em doublets} (when 
$E>E_s$) computed by using the density of states (\ref{DOSOPP}). They are
shown to be in excellent agreement with the numerical data. The splitting,
that is, 
the energy difference between the states $\ke{s}$ and $\ke{a}$ of the doublets
is represented by the third branch which is decreasing rapidly as the energy
increases. Again, the analytical result (solid curve) given by
(\ref{DeltanDWKB}) is excellent. The inset shows a comparison between this
analytical splitting and the numerical result of the numerical splitting obtained by
using Herring's formula (\ref{Herring}). 
Notice that the splitting can be now as small as
$10^{-120}$ although it has been computed with a simple FORTRAN
scheme in double precision (16 digits). 
Because the eigenvalues are of order of unity it is of
course impossible to obtain such a result by subtracting two neighboring
eigenvalues obtained by diagonalizing (\ref{Schfour}) in double precision. 
The precision is
in this case limited to $\sim 10^{-14}$. But
the way to compute the eigenvectors makes it possible inasmuch as the numerical
limit becomes not the number of digits but the smallest number representable
by the computer. This result shows that 
it is possible to compute very small splittings by using common FORTRAN
routines instead of using high precision schemes provided by
Mathematica or MAPLE for instance.

\begin{figure}[htbp]
\includegraphics[width=0.48\textwidth]{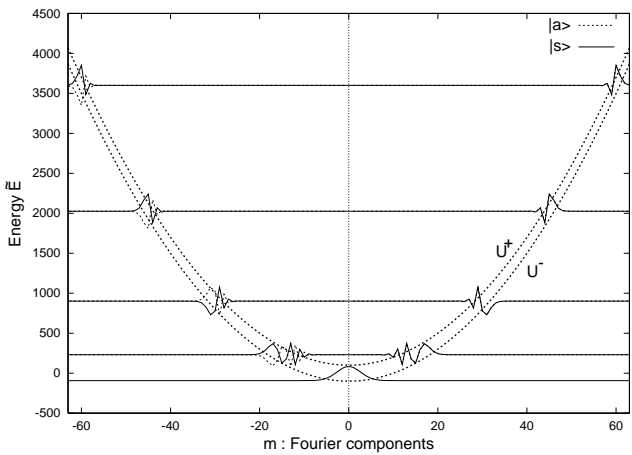}
   \caption{"Potential" curves $U^{\pm}$ used in the DWKB theory to compute
   the eigenfunctions of the SPP. Some symmetric (solid) and antisymmetric
   (dashed) eigenfunctions have been superposed to them. Their base line
   situated at their energy level allows to compare the location and the width
   of their peaks to the interval between the potential curves $U^{\pm}$. 
The value of $\alpha$ is 50. For the sake of visibility, the wave functions
   have been multiplied by a suitable scaling factor.  
   \label{DWKBwell}} 
\end{figure}

An equally important property
of the DWKB method is that it provides with
a simple and nice picture of the
basic properties of the eigenstates. Indeed, as explained in \cite{Braun93},
it is possible to associate   
a "classical" Hamiltonian defined by $H_{cl} = m^2 -2\alpha \cos \phi$ 
to the three-term recursion relation
(\ref{Schfour}), where
$m$ and $\phi$ correspond to the conjugated "coordinate" and
"momentum". This definition is rather natural as
$\hat{\phi} \equiv -i\partial/\partial m$ is precisely the Fourier
representation of the angular variable $x$. Interpreting $\phi$ as a
momentum, the expression of $H_{cl}$ shows that its classical motion is
confined between two "potential" curves defined by $U^{\pm}(m)=m^2 \pm
2\alpha$ (see Fig. (\ref{DWKBwell})). 
The idea is then to use this fact to compute the DWKB solution related
to this "classical" Hamiltonian. The internal region defined by the two
parabolas $U^{\pm}(m)$ represents the classical allowed region whereas the
regions outside are forbidden. The eigenfunctions are thus localized on the
allowed region while they decay exponentially (or faster) 
in the forbidden ones. This 
provides a natural way to express their location and their extension on the 
Fourier lattice according to their energy. We find indeed that at sufficiently
high energy (far above the separatrix) the eigenfunction of energy $E \sim
m^2/2+\alpha$ is localized around $m$ and $-m$ whatever its parity.
Its localization length around these two centers is roughly given by
$L(E) =\sqrt{\tilde{E}+2\alpha}-\sqrt{\tilde{E}-2\alpha} \sim
E_s/\sqrt{2(E-\alpha)}$ as $E \gg E_s=2\alpha$. The localization length in the
Fourier space thus decreases like the square root of the energy.

\subsubsection{Density and number of states}

\begin{figure}[htbp] 
   \includegraphics[width=0.48\textwidth]{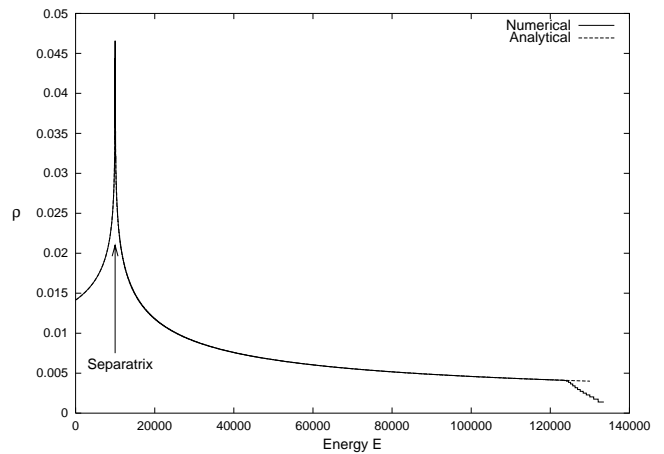}
   \caption{Density of states for a pendulum with parameter
   $\alpha=5.\,10^3$. Numerical data have been obtained from (\ref{Hfour1})
   for a matrix 
   truncated to 1001 Fourier components ($-500 \leq m \leq 500$). Comparison
   with analytical expression (\ref{DOSOPP}) 
   shows an excellent agreement until
   truncation errors become important ($E \sim 1.25\, 10^5$).
   \label{SDalp5e3}} 
\end{figure}

We end this section devoted to the properties of the single pendulum problem
 by giving the expressions of the number $n(E)$ and the density
$\rho(E)$ of states computed by the mean of Weyl formula:
\begin{equation}
\rho(E) = \frac{1}{2\pi} \int \delta (E-H(x,p))\, dx dp .
\end{equation}
It yields
\begin{eqnarray} \label{DOSOPP}
\rho(E) &=& \frac{2\sqrt{2}}{\pi\sqrt{E}} k {\cal K}(k)\ \
\ ,\ E < 2\alpha \ ;\ k=\sqrt{\frac{E}{2\alpha}} \nonumber \\
&=& \frac{2\sqrt{2}}{\pi\sqrt{E}} {\cal K}(k)\ \ \ \ \,
\ ,\ E > 2\alpha \ ;\ k=\sqrt{\frac{2\alpha}{E}}
\end{eqnarray}    
and by integrating
\begin{eqnarray} \label{nEOPP}
n(E) &=& \frac{8\sqrt{\alpha}}{\pi}  \left({\cal E}(k) - k^{\prime 2} {\cal
K}(k)\right) 
\ ,\ E \leq 2\alpha \nonumber \\
&=& \frac{4\sqrt{2E}}{\pi} {\cal E}(k)
\ ,\ E \geq 2\alpha
\end{eqnarray}
with the same definition of $k$ than above.
In these expressions ${\cal E}(k)$ and ${\cal K}(k)$ denote the usual complete
elliptic integrals of the first and second kind (see e.g. \cite{Abr}).\\
The above expressions of $\rho(E)$ show that the
density of states develops a logarithmic 
divergency close to the separatrix. In this limit, its expression
reads $\rho(E) \sim \ln(16E/|E-E_s|)/(\pi\sqrt{\alpha})$. 
This phenomenon is known as a Van Hove singularity \cite{Marder00} 
and has also been discussed for the case of the
quantum dimer \cite{Aubry96}. The graph (Fig. \ref{SDalp5e3})
shows the excellent
agreement between the density of states (DOS) computed by the mean of 
Weyl's formula and the DOS computed numerically by diagonalization of
(\ref{Schfour}).     

Using (\ref{nEOPP}), we may now obtain the splitting $\Delta E$ 
as a function of the doublet energy $E$:
\begin{equation} \label{DEe}
{\Delta E}(E) \approx \frac{\alpha^{\frac{4\sqrt{2E}}{\pi} {\cal
E}\left(\sqrt{\frac{2\alpha}{E}}\right)}}  
{\left[\Gamma
\left(\frac{4\sqrt{2E}}{\pi} {\cal
E}\left(\sqrt{\frac{2\alpha}{E}}\right)\right)\right]^2} \ \ \ (E \rightarrow
\infty) . 
\end{equation}        

\section{Two coupled pendula}

The Hamiltonian of the two coupled pendula is given by
\begin{eqnarray} \label{H2pend}
H &=& H_1 + H_2 + H_{\rm int} \nonumber \\
&=& \sum_{i=1}^{2} \left\{ \frac{p_i^2}{2}+\alpha (1-\cos x_i) \right\}
+\varepsilon (1-\cos (x_1-x_2)) \nonumber \\ 
\end{eqnarray}
where $\varepsilon > 0$ represents the coupling parameter between the
pendula.

The coupling has been chosen to be periodic in $(x_1-x_2)$ 
to allow solutions where one pendulum is oscillating whereas the
other one is rotating. This condition is of course essential to obtain
rotobreather type solutions. Notice finally that such a Hamiltonian has 
already been used to describe classical nonlinear rotating modes in a chain of
coupled pendula \cite{Takeno97}.  
   
The Hamiltonian (\ref{H2pend}) possesses two different symmetries. One is
related to the exchange of the coordinates of the two pendula $\{ {\cal S}_p
: (x_1,p_1) \leftrightarrow (x_2,p_2) \}$ and the other to the reversal 
of the
global momentum of the system, $\{ {\cal S}_m : (p_1+p_2) \leftrightarrow
-(p_1+p_2) \}$. The latter represents the generalization of the
momentum reversal symmetry already observed in the single pendulum system. 

\subsection{Classical rotobreather} \label{ClassicRoto}

\begin{figure*}[htbp]
\centering
   \includegraphics[width=0.45\textwidth]{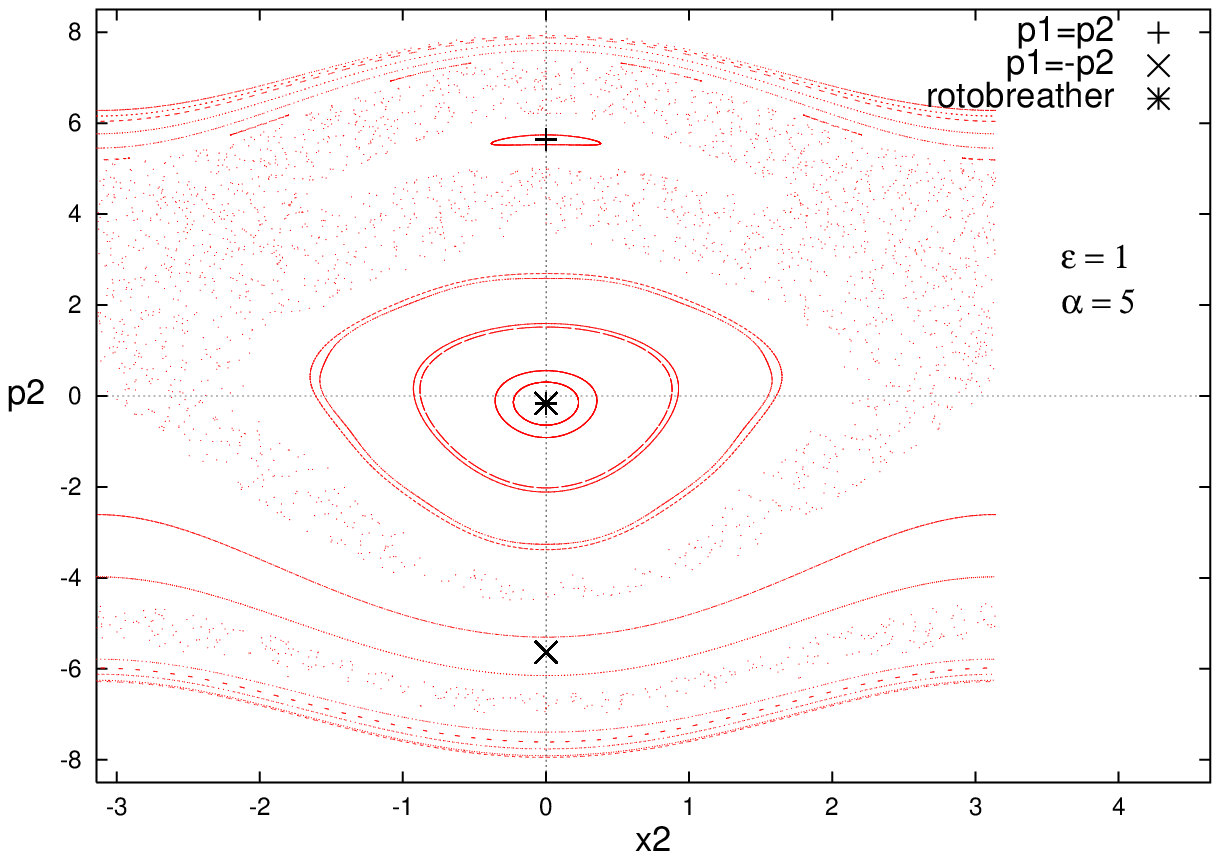} 
\hfill
   \includegraphics[width=0.45\textwidth]{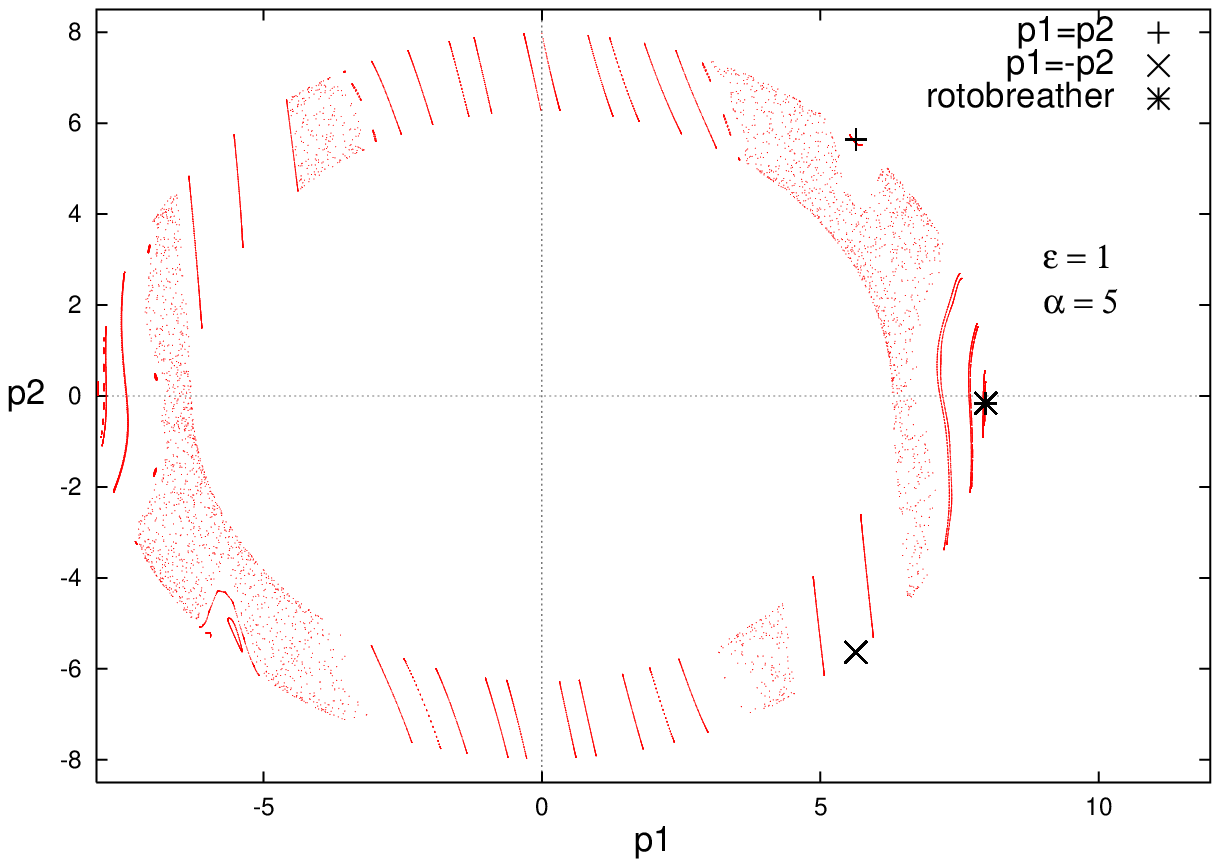}
   \caption{Poincar\'e sections of the two coupled pendula system in the
   phase space of the second pendulum (left) and in the momenta space
   (right). The conditions of the maps are $x_1=0$, $p_1>0$ (left) and 
$x_1=0$, $x_2>0$ (right). The energy $E = 31.77$. 
In-phase (cross), out-of-phase (times) and 
rotobreather (star) trajectories are represented. 
   \label{Poincpp}} 
\end{figure*}

Although the system is invariant with respect to the permutation of the
coordinates, some solutions of the equations of motion are not. Indeed, there
exist exact solutions of the Hamilton equations derived from (\ref{H2pend}),
which consist of two pendula oscillating at the same frequency but with
different amplitudes. It is also possible to
obtain solutions where one pendulum is oscillating whereas the other one is
rotating, both again evolving at the same frequency. Hereafter, both kinds of 
solutions will be respectively referred to as "breather" and "rotobreather" 
solutions. Of course, because of the size of our system, the exponential
spatial decay property of usual breathers becomes 
meaningless and the classical breather
type solution refers only to exact time-periodic solutions which break the
permutation symmetry ${\cal S}_p$.\\
As Poincar\'e sections (PS) are known to be a useful tool in describing
the behavior of non integrable dynamical systems, we will use them to locate
the orbits of the classical rotobreathers at a given energy. This energy has 
to be larger than the separatrix level of the SPP to allow one of the
pendula to rotate whereas the other one is at rest in the uncoupled
(anti-continuous) limit
($\varepsilon = 0$). But this energy can not be too close to the SPP
separatrix as $\varepsilon$ becomes non zero. Indeed, the
system develops a chaotic layer in its vicinity and prevents any stable 
periodic solution of the rotobreather type from existing. 
Concerning the classical rotobreather itself it is
obtained by using standard numerical methods like a Newton
scheme or a variational method (see e.g. \cite{Flachrep98},\cite{Marin96}).
The purpose of this paper is not to carry out an extensive study of the
classical system but merely to deal with its quantum counterpart.
Consequently we will 
just give here an example of a classical rotobreather. The values of the
parameters are $\alpha = 5$, $\varepsilon = 1$. The energy of the
rotobreather is $E_b \simeq 31.77$ and its period 
$T_b \simeq 0.885$. The initial
data are $p^b_1 \simeq 7.969$, $p^b_2 \simeq -0.160$, $x^b_1 = 0$ and
$x^b_2 = -5.2\, 10^{-9}$. The first pendulum is rotating whereas the second
one is librating. The graph Fig.\ref{Poincpp} represents two
Poincar\'e sections realized with the condition $x_1 = 0$ at an energy fixed to
$E_b$. The first PS is plotted in the phase space of the second pendulum. As
its trajectory is strictly periodic, the Poincar\'e section of the rotobreather
orbit simply consists of a single point. It has been represented by a "star"
symbol. Other periodic trajectories present in this system have been
represented by a "cross" and a "times". 
They respectively show the in-phase and
out-of-phase motions which are symmetric modes and thus can not be considered
as breathers. The second PS represents the same situation in the 
momenta space. Given the natural Fourier representation of the
quantum problem due to the $2\pi$-periodicity of the wave function, it
provides the ideal frame for the comparison of the classical and the quantum
situations. Notice that because of the existence of symmetries the presented
rotobreather solution is not the only one. Indeed, in the momentum
space $(p_1,p_2)$, the global momentum reversal symmetry is nothing but the
reflection symmetry with respect to the point $(p_1=0,p_2=0)$. And the
permutation symmetry is the mirror symmetry with respect to the line
$p_1=p_2$. Using these two symmetries we easily obtain the location of 
the four classical rotobreather solutions $(\pm p^b_1,\pm
p^b_2)$. Nevertheless, and despite these symmetries, if the system starts
initially on one of the classical rotobreather orbits, it remains on this
orbit for an infinite time. This prevents the classical system from
transferring the initial excitation from one pendulum to the other. We will see
that such a situation is impossible to be realized in the quantum system.    

In order to find
the quantum counterpart
 of the classical rotobreather, we will proceed exactly as
in the classical case. Starting from the anti-continuous limit 
where we know the classical rotobreather to exist, we will find the 
corresponding quantum state and study its evolution with respect to the
coupling $\varepsilon$.    
         
\subsection{2D Fourier space} \label{sec2Dfour}

The state of the two pendula system
$\psi(x_1,x_2)$ has to be $2\pi$-periodic in each of its variables as 
in the single pendulum problem. We thus
expand it as a double Fourier series
\begin{equation}
\ke{\psi} = \sum_{(m,n) \in \Z^2} \psi_{m,n} \ke{m,n} 
\end{equation}
where $\ke{m,n} = \ke{m}\otimes \ke{n}$ and $\bk{x}{n} = \exp
(inx)/\sqrt{2\pi}$.
With (\ref{H2pend}), the eigenvalue equation $H\ke{\psi} = E\ke{\psi}$ reads
\begin{eqnarray} \label{eigeq2D}
&& \tilde{E}\psi_{m,n} = (m^2+n^2)\psi_{m,n} -\alpha
(\psi_{m+1,n}+\psi_{m-1,n}) \nonumber \\
&& -\alpha (\psi_{m,n+1}+\psi_{m,n-1})-\varepsilon
(\psi_{m+1,n-1}+\psi_{m-1,n+1}) \nonumber \\
\end{eqnarray} 
where $\tilde{E} = 2(E-2\alpha-\varepsilon)$ is a shifted and rescaled
energy.\\
In this discrete Fourier space, the
symmetry operations ${\cal S}_m$ and ${\cal S}_p$ become:
$\{{\cal S}_m | (m,n) \rightarrow (-m,-n)\}$ (momentum reversal) and 
$\{{\cal S}_p | (m,n) \rightarrow (n,m)\}$ (permutation). 
As $H$ is invariant under these symmetries, it commutes
with the corresponding operators and its eigenstates gather naturally in
four different symmetry classes. We denote the four different kinds of eigenstates by $\{\ke{s},\ke{a},\ke{\bar
s},\ke{\bar a}\}$. Their symmetry properties are listed hereafter.   
\begin{center} 
\begin{tabular}{|c||c|c||c|c|}
\hline
State & \multicolumn{2}{c||}{Momentum reversal ${\cal S}_m$} &
\multicolumn{2}{c|}{Permutation ${\cal S}_p$} \\
\hline \hline
$\ke{s}$ & $S$ & $s_{m,n}=s_{-m,-n} $ & $S$ & $s_{m,n}=s_{n,m} $
\\
\hline
$\ke{a}$ & $A$ & $a_{m,n}=-a_{-m,-n} $ & $S$ & $a_{m,n}=a_{n,m}
$ \\
\hline
$\ke{\bar s}$ & $S$ & $\bar s_{m,n}=\bar s_{-m,-n}$ & $A$ &
$\bar s_{m,n}=-\bar s_{n,m}$ \\
\hline
$\ke{\bar a}$ & $A$ & $\bar a_{m,n}=-\bar a_{-m,-n} $ & $A$ &
$\bar a_{m,n}=-\bar a_{n,m} $ \\
\hline
\end{tabular}
\end{center}   
$S$ and $A$ indicate respectively that the state is symmetric or antisymmetric
with respect to the corresponding symmetry.
To evaluate numerically the eigenvalues and eigenvectors of (\ref{eigeq2D}),
we truncate the (normally infinite) system to $-N \leq m,n \leq N$, where $N$
is chosen sufficiently large to prevent any important truncation errors 
for states whose energy $\tilde{E} \ll 2N^2$. In this case, the total number
of computed eigenvalues is $(2N+1)^2$. Due to the symmetries,
we only diagonalize the sub-matrices representing the four different classes of
eigenvectors. The rank of the sub-matrices related to $\ke{s},\ke{a},\ke{\bar
a},$ and $\ke{\bar s}$ is $(N+1)^2$, $N(N+1)$,
$N(N+1)$ and $N^2$ respectively.

\subsection{The global spectrum at the anti-continuous limit ($\varepsilon
=0$)} 
 
There are two limiting cases where the eigenvalue equation (\ref{eigeq2D}) can
be solved analytically. These correspond to situations where the classical
system becomes integrable. The first one is realized 
when the coupling parameter $\varepsilon$ is
equal to zero, i.e. where the system consists of two identical 
decoupled pendula. The second one is realized when the
onsite parameter $\alpha$ becomes zero, where, by passing to the center-of-mass
representation, the system can be reduced to a
free rotor plus a decoupled pendulum. In these two limits 
the global spectrum of the system is given by the sum of
two one-particle spectra. 
Moreover, in the limit where $\varepsilon$ is equal to zero, the
two spectra are identical. This leads to a two-fold degeneracy of the main
part of the global spectrum. Nevertheless, by constructing the eigenvectors in
such a way that they belong to a given symmetry class, the eigenstates are
unambiguously defined and already represent the proper zeroth order states
suitable for any perturbation calculation in $\varepsilon$.

\subsubsection{Limit of zero onsite parameter: ($\alpha =0$)}

Let us first derive some results concerning the anti-continuous limit
($\varepsilon = 0$) when the onsite parameter is itself equal to zero ($\alpha
= 0$). In this case, the system consists of two free rotors and the 
global spectrum is straightforwardly given by
\begin{equation}
\tilde{E}_{m,n} = m^2 + n^2 .
\end{equation}
The corresponding eigenstates are given by a product of correctly symmetrized 
plane waves. Namely,
\begin{eqnarray}
\ke{s} &=& \frac{1}{2}\left(\ke{m,n}+\ke{-m,-n}+\ke{n,m}+\ke{-n,-m}\right)
\nonumber \\
\ke{a} &=& \frac{1}{2}\left(\ke{m,n}-\ke{-m,-n}+\ke{n,m}-\ke{-n,-m}\right)
\nonumber \\
\ke{\bar{s}} &=&
\frac{1}{2}\left(\ke{m,n}+\ke{-m,-n}-\ke{n,m}-\ke{-n,-m}\right) \nonumber \\
\ke{\bar{a}} &=&
\frac{1}{2}\left(\ke{m,n}-\ke{-m,-n}-\ke{n,m}+\ke{-n,-m}\right) . \nonumber  
\end{eqnarray} 

Notice that $\tilde{E}_{m,n}$ is a nonnegative integer in this limit.    
Its degeneracy can be expressed as follows (result due to Gauss, see
e.g. \cite{Brudern95}):\\ 
Let $\tilde{E} \in \N^*$ and $$g(\tilde{E}) = \# \{(m,n) \in \Z^2 :
m^2+n^2=\tilde{E} \}$$ be the degeneracy of the level $\tilde{E}$.
Let $$\tilde{E} = \prod_{{\rm prime} \, p} p^{e(p)}$$
be its prime number factorization. 
If $\forall \ p \equiv 3 \ {\rm mod}\ 4$, $e(p) \equiv 0\ {\rm mod}\ 2$
, then $\tilde{E}$ can be written as the sum of two squares, i.e. $g(\tilde{E})
\neq 0$ (Fermat). In this case (Gauss), its degeneracy is given by
\begin{equation} \label{degen2D}
g(\tilde{E}) = 4\prod_{p \equiv 1\, {\rm mod}\, 4}(e(p)+1) .
\end{equation}
Moreover, the number of states (i.e. the integrated density of states)
behaves asymptotically as
\begin{equation} \label{nEtilde}
n(\tilde{E}) = \sum_{\tilde{E}' \leq \tilde{E}} g(\tilde{E}') = \pi \tilde{E}
+ {\cal O}\left(\sqrt{\tilde{E}}\right) \ \ \ (\tilde{E} \rightarrow \infty) .
\end{equation}
\subsubsection{Switching on the onsite potential ($\alpha \neq 0$)} 
The last expression is not only valid in the limit of free rotors but also when
$\alpha$ and $\varepsilon$ are non zero. This is confirmed by the Weyl's
formula which indicates the zeroth order term of the degeneracy
$g(\tilde{E})$. 
Indeed (\cite{Cvitano01} p.497, {\bf 21.4})
\begin{equation}
g_{{\rm Weyl}}(\tilde{E}) =\frac{1}{2} \left(\frac{1}{2 \pi}
\int_{\tilde{V}(x_1,x_2)<\tilde{E}}\!\!\!\!\!\!
 dx_1\,dx_2 \right)
\end{equation}
where $-\tilde{V}(x_1,x_2)/2 = \alpha (\cos x_1 + \cos x_2)+\varepsilon
\cos(x_1-x_2)$.
The prefactor $1/2$ is coming from the rescaling of $\tilde{E}$ as compared
to $E$. As $(x_1,x_2) \in ]-\pi,\pi]^2$, when $\tilde{E}> \tilde{E}^+ =
{\rm max}\, (\tilde{V}(x_1,x_2))$, 
the integral term is constant and equal to $(2\pi)^2$. Thus $g_{{\rm
Weyl}}(\tilde{E}) = \pi$ and $n_{{\rm Weyl}}(\tilde{E}) = \pi \tilde{E} + 
{\rm const}$ when $\tilde{E}>\tilde{E}^+$. It is
possible to show by direct calculation that the constant 
term is equal to zero and that 
$\tilde{E}^+ = 2(2\alpha-\varepsilon)$ if $\alpha \geq 2\varepsilon$ or
$\tilde{E}^+ = 2(\varepsilon+\alpha^2/2\varepsilon)$ if $\alpha \leq
2\varepsilon$. As $\tilde{E}^+$ gives the "top" of the potential for any
values of the parameters $\varepsilon$ and $\alpha$, it also represents a
natural transition value in the global spectrum indicating the level at which
the influence of the potential $V$ starts to be weak. 
        
\begin{figure}[htbp]
\includegraphics[width=0.48\textwidth]{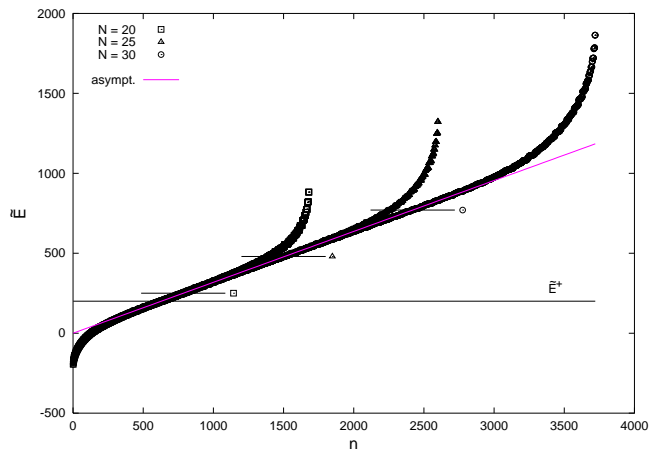}
   \caption{Spectrum of the global uncoupled ($\epsilon = 0$) system computed
   by diagonalizing (\ref{eigeq2D}) for different values of the maximal number
   of Fourier components $N$ of the wave function $\ke{\psi}$. The linear
   dimension of the Hamiltonian matrix is
   $(2N+1)^2$. $\alpha=50$. $\tilde{E}^+$ (see text) has been indicated as
   well as the thresholds where, because of truncation, the computed spectrum
   starts to be larger gapped. Thresholds are represented by straight lines ended by
   the symbol of the correponding spectrum. 
The asymptote $\tilde{E} = n/\pi$ is also shown. 
   \label{Specalp5e1eps0}} 
\end{figure}

Expression
(\ref{nEtilde}) thus provides a very useful guide for checking the energy
level at which truncation errors become important. 
Note that truncation
errors are of two types. First, because of the truncation of the whole matrix,
there exists an energy threshold for which 
some eigenvalues of the spectrum are missing. 
Indeed for $\epsilon = 0$, the global spectrum is given by the
sum of the SPP spectrum with itself. In this case, even assuming
that the numerically computed eigenvalues are exact,
the SPP spectrum ends at $\tilde{E}^{\rm SPP}_e$ because of truncation. 
Let us denote the energy of its ground state by
$\tilde{E}^{\rm SPP}_0$. Then 
the first eigenvalue to be missed is $\tilde{E}^{\rm SPP}_{e+1}+\tilde{E}^{\rm
SPP}_0$ where $\tilde{E}^{\rm SPP}_{e+1}$ is the first eigenvalue following 
$\tilde{E}^{\rm SPP}_e$. This provides with a 
threshold where the eigenvalues start to be ranked in a wrong way. Their
label becomes false and so does the computed number of states $n(\tilde{E})$.
This is shown in Fig. \ref{Specalp5e1eps0} 
which represents the spectrum of
two uncoupled pendula as computed from equation (\ref{eigeq2D}) for
different values of $N$ (recall that $-N \leq m,n \leq N$).
The second type of error induced by the
truncation is a modification of the values of the energies themselves. As we
have verified these errors increase as we reach the upper
end of the truncated spectrum 
but are nevertheless very small if we respect the threshold
indicated above.     

Although 
the degeneracy (\ref{degen2D}) has been obtained for the case
where both the onsite and the coupling parameters are zero, it provides
with useful information when considering  
the near degeneracy of $n$-uplets arising at
sufficiently high energy (that is, far above twice the energy of the
separatrix of the SPP).

As the onsite parameter $\alpha$ becomes different from zero, the first part
of the free rotor spectrum of the SPP is modified. Far below the SPP's 
separatrix, 
the quadratic spectrum $\tilde{E}^{\rm SPP}_l\!\!=\!l^2$ is replaced by
a harmonic oscillator one $\tilde{E}^{\rm SPP}_l\!\! \sim \!
2(\sqrt{\alpha} l-\alpha)$.  
The number of
states $n(\tilde{E})$ of the global uncoupled system thus becomes a quadratic
function of the energy. This explains the form of the function $\tilde{E}(n)$
observed in Fig.\ref{Specalp5e1eps0} which starts as a square root and
asymptotically becomes a linear function of $n$.

\subsection{Quantum rotobreather state}

\subsubsection{Anti-continuous limit}

\begin{figure}[htbp]
\includegraphics[width=0.48\textwidth]{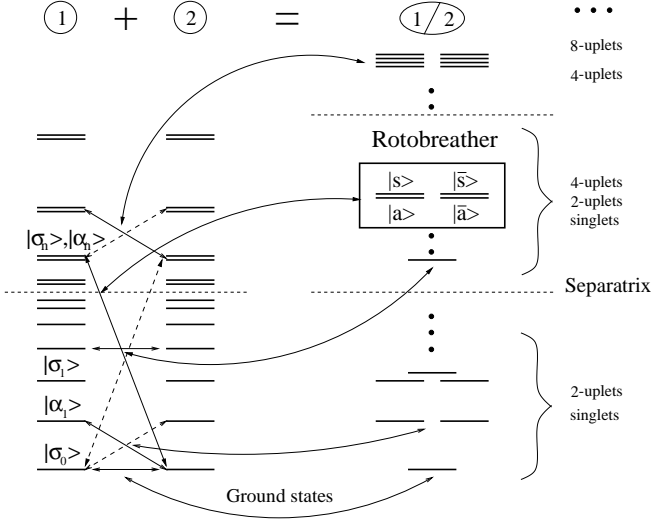}
   \caption{Schematic representation of the sum of two pendula
   spectra. Straight solid
   arrows indicate the levels to be added and dashed arrows the symmetric
   (permutation) operation. The result is indicated in the global spectrum by
   a curved arrow. The
   construction of the quantum rotobreather state $\ke{\Phi}$ is explicitely 
represented.
   \label{addpend12}} 
\end{figure}
The goal of this paper is to define and to study a quantum rotobreather
state $\ke{\Phi}$ whose properties are very similar to those of the classical
rotobreather. Consequently it is natural to look 
for a quantum state which, at the
anti-continuous limit ($\varepsilon = 0$), represents a state consisting of a
rotating pendulum and another one at rest (up to quantum fluctuations).
As the pendula are not coupled, this state is represented by the tensorial
product of the SPP's ground state $\ke{\sigma_0}$ (corresponding to the
pendulum at rest) and a superposition of two states 
($\ke{\sigma_n},\ke{\alpha_n}$) belonging 
to the same doublet $n$ of the SPP (corresponding to a rotating pendulum).
The addition of the two SPP spectra and the construction of the quantum
rotobreather $\ke{\Phi}$ is schematically depicted in Fig. \ref{addpend12}.
The states denoted by $\ke{\sigma}$ ($\ke{\alpha}$) are symmetric
(antisymmetric) with respect to the inversion of the momentum:
\begin{equation} \label{QuadPhi}
\ke{\Phi} = \frac{1}{\sqrt{2}} \ke{\sigma_0} \otimes 
(\ke{\sigma_n}+\ke{\alpha_n}) .
\end{equation}
In order to obtain the state $\ke{\Phi}$ at $\varepsilon = 0$  
as a linear combination of the correctly symmetrized eigenvectors 
listed in the section \ref{sec2Dfour}, 
we first construct it in terms of the SPP
eigenvectors. The correct symmetrization of
the tensorial products leads to the following expression of the four possible
classes of states: 

\begin{eqnarray}
\ke{s} &=& \frac{1}{\sqrt{2}} \left(\ke{\sigma} \otimes \ke{\sigma '} + 
\ke{\sigma '} \otimes \ke{\sigma}\right) \nonumber \\ 
&{\rm or}& \frac{1}{\sqrt{2}} \left(\ke{\alpha} \otimes \ke{\alpha '} + 
\ke{\alpha '} \otimes \ke{\alpha}\right) \nonumber \\
\ke{\bar{s}} &=& \frac{1}{\sqrt{2}} \left(\ke{\sigma} \otimes \ke{\sigma '} - 
\ke{\sigma '} \otimes \ke{\sigma}\right) \nonumber \\ 
&{\rm or}& \frac{1}{\sqrt{2}} \left(\ke{\alpha} \otimes \ke{\alpha '} - 
\ke{\alpha '} \otimes \ke{\alpha}\right) \nonumber \\
\ke{a} &=& \frac{1}{\sqrt{2}} \left(\ke{\sigma} \otimes \ke{\alpha} + 
\ke{\alpha} \otimes \ke{\sigma}\right) \nonumber \\
\ke{\bar{a}} &=& \frac{1}{\sqrt{2}} \left(\ke{\sigma} \otimes \ke{\alpha} - 
\ke{\alpha} \otimes \ke{\sigma}\right) 
\end{eqnarray}
where $\ke{\sigma},\ke{\sigma '}$ ($\ke{\alpha},\ke{\alpha '}$) 
represent any of the symmetric (antisymmetric) SPP eigenvectors.

Using these expressions, $\ke{\Phi}$ is readily written as
\begin{equation}
\ke{\Phi} = \frac{1}{2} \left(\ke{s}+\ke{\bar{s}}+\ke{a}+\ke{\bar{a}}\right)
\end{equation}
where $\ke{s},\ke{\bar{s}} = (\ke{\sigma_0} \otimes \ke{\sigma_n} \pm
\ke{\sigma_n} \otimes \ke{\sigma_0})/\sqrt{2}$ and $\ke{a},\ke{\bar{a}} =
(\ke{\sigma_0} \otimes \ke{\alpha_n} \pm \ke{\sigma_n} \otimes
\ke{\alpha_0})/\sqrt{2}$. 
In this anti-continuous limit, we may compute the
eigenvalues corresponding to each of the four states which contribute
to $\ke{\Phi}$.
We obtain 
\begin{equation} \label{Enquadeps0}
\tilde{E}_s = \tilde{E}_{\bar{s}} = \tilde{E}_{\sigma_0} + \tilde{E}_{\sigma_n}
\ \ ;\ \ 
\tilde{E}_a = \tilde{E}_{\bar{a}} = \tilde{E}_{\sigma_0} +
\tilde{E}_{\alpha_n} . 
\end{equation}
As the expression (\ref{Enquadeps0}) shows, the absence of any coupling
between the pendula is responsible for the true degeneracy of the states 
whose parity is the same with respect to the momentum reversal symmetry .
This follows from the fact that 
the global spectrum is given by the sum of the SPP spectrum with itself
in the anti-continuous limit (cf. Fig. \ref{addpend12}). 
Thus the energy level $\tilde{E}_{\sigma_0} +
\tilde{E}_{\sigma_n}$ 
is obtained in this limit either by adding the energy of the ground
state $\ke{\sigma_0}$ of the first pendulum to the energy of the symmetric
state $\ke{\sigma_n}$ of the $n^{\rm th}$ doublet of the second pendulum, or
by adding the energy of the ground
state of the second pendulum to the energy of the symmetric
state of the first one. 

We have already shown (see \ref{DEn}) 
that the presence of an onsite potential 
($\alpha \neq 0$) lifts the degeneracy of the symmetric and antisymmetric
states belonging to the same doublet of the SPP. We may thus compute the
splitting between the two pairs of degenerated states $(\ke{s},\ke{\bar{s}})$
and ($\ke{a},\ke{\bar{a}}$):
\begin{equation} \label{DEnquadeps0}
\delta_n = \tilde{E}_s - \tilde{E}_{a} = 2 \frac{\alpha^{2n}}{(2n-1)!^2} .
\end{equation}
For sufficiently large $n$ this splitting is extremely small and thus the four
states making up $\ke{\Phi}$ form a {\em quadruplet} of nearly degenerated
eigenenergies which has been represented in Fig. \ref{addpend12}. 
Notice finally that the inverse of this splitting is directly
related to the time taken by the system whose initial state is $\ke{\Phi}$ 
to reverse its total momentum. As one of the pendula is at rest (up to
quantum fluctuations), this means that the pendulum initially 
rotating in a
given sense tunnels into the state which corresponds to the opposite 
rotation sense on a time scale $\tau_p \sim 1/\delta_n$. 
This effect is purely quantum.

\subsubsection{Nonzero coupling $\varepsilon$}

As soon as the coupling parameter $\varepsilon$ is nonzero the degeneracy of
the states $(\ke{s},\ke{\bar{s}})$ and ($\ke{a},\ke{\bar{a}}$) of the
quadruplet is lifted. 
Here we are interested in the
computation of the corresponding eigenvalues in order to find the splittings
which determine
the evolution of quantities such as individual
energies or momenta of the pendula. If the splittings
between the states making up the quantum breather become of the order of
the mean level spacing of the spectrum, we may conclude that the breather
solution is lost \cite{FlachFle97}. 

One possible source of a dramatic increase of the splittings
can be a strong overlap of the quantum state with the chaotic layer of
the classical system (see e.g. \cite{Bohigas93} for a general review of the
manifestations of classical phase space structures in quantum mechanics). 
This overlap is generally computed by the mean of a 
Husimi distribution which is one of the possible phase-space representations
of a quantum state (see e.g. \cite{Lee95}). 
This distribution is then superposed to the
corresponding Poincar\'e section of the classical system which allows to
compute the overlap. This method has for instance been used in the case of a
driven bistable system \cite{Utermann94}. At the same time
these studies have shown that doublet states 
overlapping up to  70\% with the chaotic layer may still possess a small
splitting. To avoid misinterpretations,
here we will compare the Fourier components of the quantum
rotobreather $\ke{\Phi}$, with the
Poincar\'e section of the classical system in the momenta space. The main
result will be
that the phase space locations of the classical and the quantum rotobreathers
are roughly the same. 

This direct comparison is possible due to the results
obtained in \cite{Torres98} showing that the Husimi distribution of the
pendulum problem can be found analytically. 
It follows that for the
pendulum potential, the discrete Fourier representation of the eigenfunctions 
and their Husimi distributions restricted to the momentum space differ
insignificantly.  

\subsubsection{Tunneling of the $\ke{\Phi}$ state}

In this section we derive the expressions of some relevant quantities 
which allow us to follow the time evolution of the initial state $\ke{\Phi}=
(\ke{s}+\ke{\bar{s}}+\ke{a}+\ke{\bar{a}})/2$. This state is
formed at $\varepsilon \neq 0$ by the eigenstates which belong to the
quadruplet of states.
These states in turn are defined  
by the tensorial product of the
ground-state of the SPP spectrum and one of its doublets at the anti-continuous
limit. Because $\ke{\Phi}$
is not an eigenstate it evolves in time. In order to visualize this evolution
and because we are working in the 2D discrete Fourier space (2DFS), 
we may compute
the time evolution of its momentum. This is done by computing the two
functions 
$\langle \Delta P \rangle = \moy{\Phi}{\hat{P}_1-\hat{P}_2}{\Phi}$ and
$\langle P \rangle = \moy{\Phi}{\hat{P}_1+\hat{P}_2}{\Phi}$ 
where $\hat{P}_1$ and $\hat{P}_2$ are
respectively the momenta operators of the pendula 1 and 2. If the total
momentum of the system were conserved (as in the integrable limit $\alpha=0$),
the difference of the momenta $\langle \Delta
P \rangle$ would exactly 
represent the transfer of momentum between the pendula. 
This is not the case for a nonzero value of $\alpha$. 
Although the total momentum $\langle P \rangle$ is in
general not conserved, for not too small values of $\varepsilon$
it evolves very slowly as compared to $\langle \Delta P \rangle$. We
thus may consider that the latter represents 
the transfer of the excitation from pendulum 
to pendulum with a good accuracy.\\
Another quantity of interest is the difference between the
individual (or onsite) energies $\langle \Delta H \rangle =
\moy{\Phi}{\hat{H}_1-\hat{H}_2}{\Phi}$ where $\hat{H}_i = \hat{P}^2_i/2+
\alpha (1-\cos \hat{X}_i)$. As the total energy of the system is conserved,
$\langle \Delta H \rangle$ measures the transfer of energy between the
pendula. \\
In order to give a simple expression of the above quantities let us use 
the two symmetries of permutation and momentum reversal. We denote their
respective operators by ${\cal P}$ and ${\cal M}$. 
It is easily shown that they are unitary: 
$ {\cal P}^{\dagger}{\cal P} = {\cal M}^{\dagger}{\cal M} = \One \ $ 
(see e.g. \cite{Cohen77}). Moreover, the following relations hold:
\begin{eqnarray}
{\cal M}^{\dagger}\hat{P}_i{\cal M} = - \hat{P}_i \ \ \ &;& 
\ \ \ {\cal P}^{\dagger}\hat{P}_i{\cal P} =  \hat{P}_j \\
{\cal M}^{\dagger}\hat{H}_i{\cal M} =  \hat{H}_i \ \ \ &;& 
\ \ \ {\cal P}^{\dagger}\hat{H}_i{\cal P} =  \hat{H}_j \\
{\cal M}^{\dagger}\hat{H}_{\rm int}{\cal M} =  \hat{H}_{\rm int} \ \ \ &;& 
\ \ \ {\cal P}^{\dagger}\hat{H}_{\rm int}{\cal P} =  \hat{H}_{\rm int}
\end{eqnarray}	
where $(i,j) \in (1,2), i \neq j$ and where $\hat{H}_{\rm int} = \varepsilon
(1-\cos (\hat{X}_1-\hat{X}_2))$ is the
interaction energy term.\\
Using the fact that the Hilbert space $\cal E$ associated to the coupled
pendula problem can be written as the direct sum ${\cal E} = {\cal E}_s
\oplus  {\cal E}_{\bar s} \oplus {\cal E}_a \oplus {\cal E}_{\bar a}$, 
any operator may be represented by a $4\times 4$
block-matrix in the basis $(\{\ke{s_i}\},\{\ke{\bar
s_j}\},\{\ke{a_k}\},\{\ke{\bar a_l}\})$ formed by all the states belonging to
${\cal E}_s$ first, then all the 
states belonging to ${\cal E}_{\bar s}$, etc... in the order described above.  
From the preceeding relations and by using the symmetry properties of the
eigenstates, it follows 
that the relevant operators for our study have the form
     
\begin{equation} \label{Pmatrices}
\hat{P}_{1,2} = \left(
\begin{array}{cccc}
   0 & 0 & P_{as} & \pm P_{{\bar a}s} \\
   0 & 0 & \pm P_{a{\bar s}} & P_{{\bar a}{\bar s}} \\
   P_{sa} & \pm P_{{\bar s}a}& 0 &  0 \\ 
   \pm P_{s{\bar a}} & P_{{\bar s}{\bar a}} & 0 &  0
\end{array}
\right) 
\end{equation}
and
\begin{equation} \label{Hmatrices} 
\hat{H}_{1,2} = \left(
\begin{array}{cccc}
   H_{ss} & \pm H_{{\bar s}s} & 0 & 0 \\
   \pm H_{s{\bar s}} &H_{{\bar s}{\bar s}}  & 0 & 0 \\
   0 & 0 & H_{aa} & \pm H_{{\bar a}a}  \\
   0 & 0 & \pm H_{a{\bar a}} &H_{{\bar a}{\bar a}} 
\end{array}
\right) .
\end{equation}       
We have used the condensed notation $O_{\mu \nu} =
\moy{\mu}{\hat{O}_1}{\nu}$ which symbolically represents all the matrix
elements of the observable $\hat{O}_1$ between states of two given 
subspaces ${\cal E}_\mu$ and ${\cal E}_\nu$. 
The sign $\pm$ is $+$ for 1 and $-$ for
$2$. In this basis, the operator $\hat{H}_{\rm
int}$ is block-diagonal.\\ 
Using the time evolution of the initial state $\ke{\Phi}$
\begin{equation}
\ke{\Phi (t)} = \frac{1}{2} \sum_{\mu} e^{-iE_{\mu}t}\ke{\mu} \ \ \
\text{where}\ \ \ \mu \in \{s,{\bar s},a,{\bar a}\}  
\end{equation}      
we finally obtain
\begin{eqnarray} 
\langle \Delta P \rangle &=& \moy{s}{\hat{P}_1}{\bar a}\cos (E_s-E_{\bar a})t 
\nonumber \\
&+& \moy{{\bar s}}{\hat{P}_1}{a}\cos (E_{\bar s}-E_a)t \label{DP} \\
\langle  P \rangle &=& \moy{s}{\hat{P}_1}{a}\cos (E_s-E_a)t \nonumber \\
&+& \moy{{\bar s}}{\hat{P}_1}{\bar a}\cos (E_{\bar s}-E_{\bar a})t \label{P} 
\\
\langle \Delta H \rangle &=& \moy{s}{\hat{H}_1}{\bar s}\cos (E_s-E_{\bar s})t
\nonumber \\ 
&+& \moy{a}{\hat{H}_1}{{\bar a}}\cos (E_a-E_{\bar a})t . \label{DH}
\end{eqnarray}
Notice that the
energy differences 
occurring in $\langle \Delta H \rangle$ are zero at the anti-continuous limit. No transfer of
energy between the pendula occurs due to the fact that
they are decoupled. The excitation (and thus the energy) is entirely conserved
on its initial site. In this limit we know (in leading order) the value of the
splittings $E_s-E_a=E_{\bar s}-E_{\bar a}=\alpha^{2n}/(2n-1)!^2$ and thus the
time taken by the system to reverse its initial momentum: $\tau_p = \pi
(2n-1)!^2/ \alpha^{2n}$ where $n$ labels the doublet of the rotating pendulum.
These splittings also occur in $\langle \Delta P \rangle$.  
Thus for small $\varepsilon$ 
we can not assign the meaning of a transfer of momentum between the
pendula to the time evolution of $\langle \Delta P \rangle$. 

In the other integrable limit where $\varepsilon \neq 0$ but
$\alpha = 0$, it is also possible to compute the exact spectrum of the
system. Notice that the total momentum of the system is strictly
conserved in this case. The absence of the
onsite potential is also responsible for the degeneracy of the states of
different parity concerning the momentum reversal symmetry. We thus have
$E_s=E_a$ and $E_{\bar s}=E_{\bar a}$. It is possible to show (see
appendix A) that
the splitting $E_s-E_{\bar s}=(a_n(-2\varepsilon)-b_n(-2\varepsilon))/4$ which
in leading order gives $|E_s-E_{\bar s}|=2(\varepsilon /2)^n/(n-1)!^2$ where
$n$ is the label of the doublet of the initially rotating pendulum. The time
taken by the system to transfer its excitation  
from one site to the second one is $\tau_e
= \pi (n-1)!^2/2(\varepsilon /2)^n$ in this limit.
By comparing $\tau_e$ with $\tau_p$ we observe that for large enough $n$,
$\tau_p \gg \tau_e$ holds, 
regardless the values of $\varepsilon$ and $\alpha$. 
This indicates that in general transfer of energy (and momentum)
from site to site will be a much faster process than the process
of total momentum reversal.

When $\alpha$ and $\varepsilon$ are both nonzero, the different splittings
have to be calculated numerically by following the evolution of the quadruplet
as a function of the coupling $\varepsilon$. This is the purpose of the
following section.

\subsubsection{Following the quadruplet}

\begin{figure}[htbp]
\includegraphics[width=0.48\textwidth]{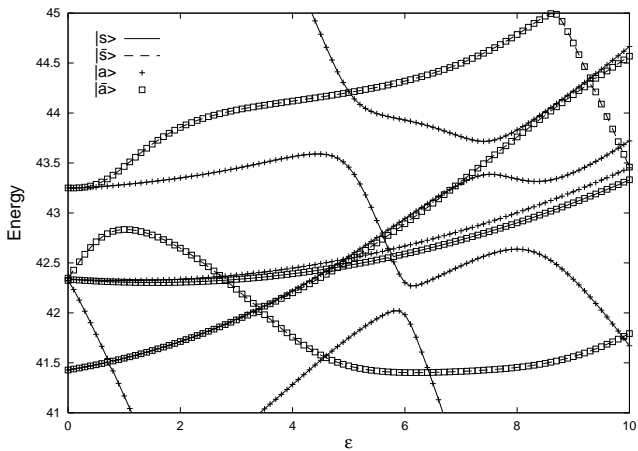}
   \caption{Evolution of a part of the spectrum as a function of the
   coupling $\varepsilon$. The onsite parameter is $\alpha=5$. Energies are
   rescaled according to $\tilde{E}=2(E-2\alpha-\varepsilon)$. The quadruplet
   under investigation starts at an energy $\tilde{E} \simeq 41.43$. It
   corresponds to a state formed of the ground-state of the SPP and the doublet
   $n=7$.  
   \label{Detevolspeceps}} 
\end{figure}

In Fig. \ref{Detevolspeceps}, we have plotted the evolution of a part
of the global spectrum of the coupled pendula as a function of the coupling
$\varepsilon$. The onsite parameter is $\alpha=5$. The energies of the
different classes of states have been represented by different
lines and symbols. 
The quadruplet which
represents the levels of the states
$\{\ke{s},\ke{\bar{s}},\ke{a},\ke{\bar{a}}\}$ from which $\ke{\Phi}$ is
formed, has an energy $\tilde{E} \simeq 41.43$ at $\varepsilon=0$. 
This corresponds to a
quadruplet made of the ground-state of the SPP and
the doublet $n=7$ at the anti-continuous limit. 
Indeed, by using (\ref{Lowen}) and (\ref{Endoub}), we obtain
$\tilde{E}^d_7 \simeq 49+50/195$ and $E_0 \simeq \sqrt(5)-1/16 \Rightarrow  
\tilde{E}_0 \simeq \sqrt(5)-1/16-10$. The sum of these energies gives
$\tilde{E} \simeq 41.43$. At $\varepsilon = 1$ the quadruplet energy is
$\tilde{E} \simeq 41.54$. This corresponds to $E \simeq 31.77$ which precisely 
represents the energy of the classical rotobreather presented in the section 
\ref{ClassicRoto}.
   
At slightly larger energy in the
spectrum we observe an octuplet. It consists of states mixing two neighboring
doublets, namely $n=4$ and $n'=5$.
Its energy is around $\tilde{E} \simeq
42.3$ which can be computed also by using (\ref{Endoub}). 
The reason for the occurence of this octuplet is depicted in
Fig. \ref{addpend12}. 
By combining the two states of the 
doublet $n$ with those of $n'$ we obtain four different (although nearly
degenerated) levels of energy. 
But there are two ways to attribute these doublets to the pendula 1 and 2
(represented by the solid and dashed arrows in Fig. \ref{addpend12}). We
thus obtain two identical quadruplets which yield an octuplet.
Finally, the visible cluster of levels at $\tilde{E} \approx 43.25$ 
corresponds to a quadruplet made
of the combination of a single symmetric state (the last one situated below
the separatrix $\tilde{E}_{\rm sep}=2\alpha=10$) with the doublet $n=6$.

From a general point of view, the evolution of the energy levels of a spectrum
as a function of a given parameter (here $\varepsilon$) can be compared to
the time evolution of a gas of particles obeying dynamical laws of the
Calogero-Moser system type (\cite{Pech83}-\cite{Haake01}). This
system is Hamiltonian and the interaction between particles (eigenvalues) is
strongly repulsive at short distances. This gives rise to avoided
crossings. Nevertheless, in the presence of symmetries, the different parity
sectors decouple from each other and thus, do not
interact \cite{Wilki87}. 
As a result, the four sectors $\{s,{\bar s},a,{\bar a}\}$
evolve individually according to the Calogero-Moser dynamics but do not 
interact with each other. Thus they may cross. 
Such a crossing is observed in
Fig. \ref{Detevolspeceps} in the vicinity of the point $(\varepsilon
=5,\tilde{E}=44)$.\\
Fig. \ref{Detevolspeceps} also shows the pairing of eigenvalues whose
momentum related parities are complementary. For instance
eigenvalues of the $s$ sector cluster with eigenvalues of the $a$ or
${\bar a}$ sectors. They are indeed quasi-indistinguishable. The reason is
that the influence of the onsite potential becomes
weak and thus the corresponding splittings small as the energy becomes 
sufficiently high  .

\begin{figure}[htbp]
\includegraphics[width=0.48\textwidth]{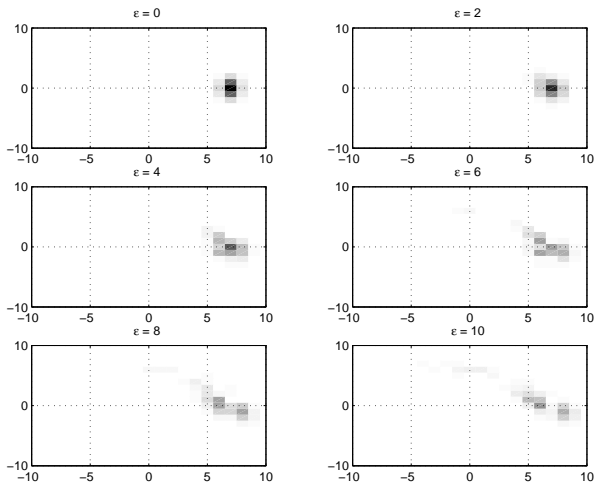}
\vskip 0.5cm
   \caption{Evolution of $\ke{\Phi} \ \ (|\Phi_{m,n}|^2)$ as the coupling 
   strength $\varepsilon$ increases. Each picture represents the 2D Fourier 
   space, $p_1 (m)$ along the $x$ axis and $p_2 (n)$ along the $y$ axis. 
The value of each
   component $|\Phi_{m,n}|^2$ is represented by its gray level from white (0)
   to black $(\sim 0.3)$ .
   \label{EvolPhiepsvar}} 
\end{figure}

By following the quadruplet which starts at $\tilde{E} \simeq 41.43$, we
observe 
that it first survives an avoided crossing with one $(\ke{s},\ke{a})$ pair
($\varepsilon \sim 0.7$). Then it survives again another one with two
$(\ke{{\bar s}},\ke{{\bar a}})$ states and starts to split into two pairs
$(\ke{s},\ke{a})$ and $(\ke{{\bar s}},\ke{{\bar a}})$ after a collision with a
quasi-quadruplet which originates from the octuplet at $\tilde{E} \simeq
42.3$. Finally, in the vicinity of $\varepsilon = 7.5$, the $(\ke{{\bar
s}},\ke{{\bar a}})$ pair seems to join a new $(\ke{s},\ke{a})$ pair but
remains clearly separated from it.
The fact that pairs of permutationally related eigenvalues may survive avoided
crossings is already known and may for instance be observed in the case of
the trimer problem (\cite{FlachFle97},\cite{FlachFle01}).

In order to see the progressive evolution of our initial state $\ke{\Phi}$ as
the coupling increases, we have plotted snapshots of its evolution for 
different values of $\varepsilon$ (Fig. \ref{EvolPhiepsvar}). Each picture
represents the 2D discrete Fourier space. The $x$-axis represents the
coordinate $m \in \Z$ related to the momentum of the first pendulum $p_1$
and the $y$-axis, $n \in \Z$ related to $p_2$. The $\ke{\Phi}$ state has been
computed on a square lattice of $41 \times 41$ Fourier components. For 
clarity only a $21 \times 21$ lattice has been displayed. We 
observe that the state $\ke{\Phi}$ remains almost unchanged
for $0 \le \varepsilon \le 2$. 
The excitation is well localized in the 2D Fourier
space. The average momentum $\langle \hat{P}_1\rangle =
\sum_{m,n}m|\Phi_{m,n}|^2$, which defines the projection of the excitation
"center" on the $x$-axis, is large and makes the first pendulum "rotating"
whereas $\langle \hat{P}_2\rangle = \sum_{m,n} n |\Phi_{m,n}|^2$ is small and
makes the second pendulum "oscillating". The quotes indicate that this
correspondence
refers to an interpretation in terms of the classical 
system. As $\varepsilon$ becomes larger than $4$, the
state starts to spread on the lattice and does not correspond to a "coherent"
excitation anymore. A plot of the Poincar\'e section of the classical system
at $\varepsilon = 6$ with an energy corresponding to the one 
of $\ke{\Phi}$ has shown the
allowed region of the Fourier space to be chaotic (except in the vicinity
of the in-phase and out-of-phase motion and in a tiny region where the
classical rotobreather still exists). This means that any small perturbation of
the initial conditions of the classical rotobreather leads to a chaotic
trajectory. We verify in this particular example that the strong overlap of
the chaotic sea with the quantum state leads to large splittings of tunneling
pairs (here $(\ke{s},\ke{{\bar s}})$ and $(\ke{a},\ke{{\bar a}})$). 

\subsubsection{Evolution of the splittings with $\varepsilon$}

\begin{figure}[htbp]
\includegraphics[width=0.48\textwidth]{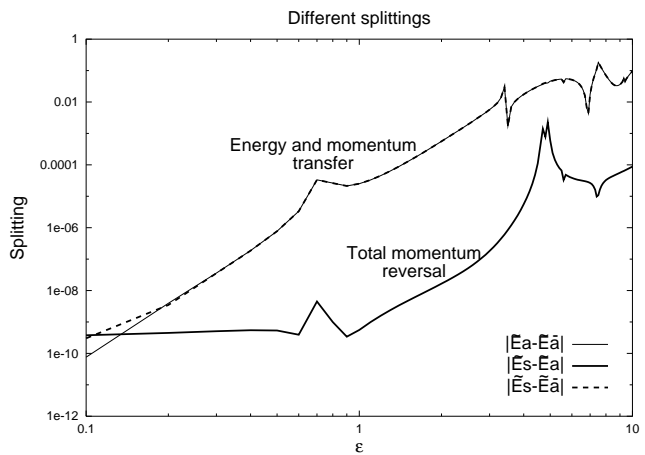}
   \caption{Dependence  of different splittings of the
   quadruplet whose energy 
   at $\varepsilon=0$ is $\tilde{E} \simeq 41.43$ 
   on $\varepsilon$. 
   Only three of them have been displayed, each being associated with a given
   tunneling process. The splittings are
   given in terms of the rescaled energy $\Delta \tilde{E}=2 \Delta E$.   
   \label{splittE42}} 
\end{figure}

To provide quantitative results concerning the behavior of the 
different splittings involved in the expressions (\ref{DP}) 
as $ \varepsilon$ increases, we have
computed them over the same range of values. The
result is presented in Fig. \ref{splittE42}. It always concerns 
the state $\ke{\Phi}$ obtained from the quadruplet of
preceeding section. Notice that for the sake of clarity, only three of the six
possible splittings have been displayed on the figure. The omitted splittings
behave very similarly according to the symmetry
properties of the involved states. Each of the displayed splittings 
corresponds to a
given tunneling process as specified in the figure. After 
a crossing value $\varepsilon \simeq 0.15$,
the graph basically consists of two curves. One is related to the momentum 
or energy transfer between the pendula and another one is related to
the reversal of the total momentum of the system. This means that the upper
curve is related to the transfer of the excitation from a general point of
view (energy or momentum). The lower curve represents a process (the global
momentum reversal) which is not associated to any kind of excitation transfer
between the pendula but only to a global modification of the system. It has
no relation with the tunneling of the quantum rotobreather from site to site.

Note that at the anti-continuous limit, the energies of the states
$\ke{a}$ ($\ke{s}$) and $\ke{\bar a}$ ($\ke{\bar s}$) are identical. 
The splitting which corresponds to the energy transfer
($|\tilde{E}_a-\tilde{E}_{\bar a}|$ in the figure) is thus zero. 
This explains the
behavior of the corresponding (thin solid) curve, in logarithmic scale, which
decreases regularly with $\varepsilon$. 
Moreover, due to this
degeneracy, all the remaining splittings are equal and given by 
(\ref{DEnquadeps0}). For the $\ke{\Phi}$ state we obtain
$\delta_7 \simeq 3.15\,10^{-10}$ which corresponds to the 
saturation value obtained numerically for $|\tilde{E}_s-\tilde{E}_a|$ and 
$|\tilde{E}_s-\tilde{E}_{\bar a}|$ (thick solid and dashed lines).
  
The splittings depend smoothly on $\varepsilon$  
except for values of the
coupling where crossings and avoided crossings take place (see Fig. 
\ref{Detevolspeceps}).
The first crossing involves the quadruplet states
$\{\ke{s},\ke{a},\ke{\bar s},\ke{\bar a}\}$ and a
($\ke{s'},\ke{a'}$) pair at $\varepsilon \simeq 0.73$. Avoided crossings occur
between $(\ke{s},\ke{s'})$ and $(\ke{a},\ke{a'})$ pairs and are responsible
for the peaks appearing in the three splittings. 
The case of the second crossing ($\varepsilon
\simeq 3.30$) is slightly
different. This avoided crossing concerns the 
$(\ke{\bar s},\ke{\bar s'})$ and $(\ke{\bar a},\ke{\bar a'})$ pairs. 
The splitting $|\tilde{E}_s-\tilde{E}_a|$ behaves smoothly 
whereas the two others exhibit spikes due to the
collision. This follows from the fact that
 $|\tilde{E}_s-\tilde{E}_a|$ does not contain any
contribution of states interacting with the colliding pair. 
Note that the spikes observed in such crossings may have
different forms according to the exact realization of the collision and the
specific splitting under consideration \cite{FlachFle01}. 

Finally, when the coupling parameter becomes larger than  5, the
splittings corresponding to the transfer of the excitation between the
pendula start to be of the same order as the mean spacing in the spectrum
($1/\pi$ here as shown in (\ref{nEtilde})). This is the signature of the
disappearance of the quantum rotobreather since no quantum state is able to
keep the energy on a given pendulum during a  
sufficiently long time at the considered energy level.

\begin{figure}[htbp]
\includegraphics[width=0.48\textwidth]{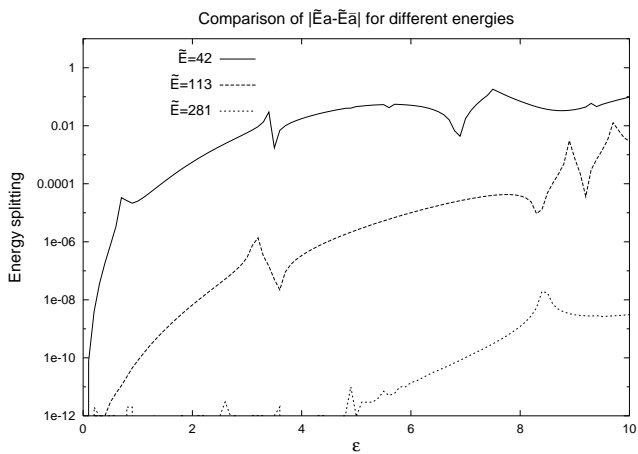}
\vskip 0.5cm
   \caption{Dependence  of the splitting governing the energy
   tunneling on $\varepsilon$ for different quadruplet energies. 
   The three curves correspond to
   quadruplets starting from a ground state and a doublet at
   $\varepsilon=0$. The respective 
labels are $m=7,\ 11\ \text{and }\ 17$. The corresponding energies at
   $\varepsilon=0$ are $\tilde{E} = 41.43,\ 113.3\ \text{and }\ 281.2$.     
   \label{splittE42E113E281}} 
\end{figure}

We conclude this section by discussing  
the dependence of the
splitting governing the energy transfer on the 
quadruplet energy (Fig. \ref{splittE42E113E281}). 
As expected the splittings decrease with increasing
quadruplet energy. This can be
qualitatively understood by referring to the analytical expression of the
splitting at $\alpha = 0$. Indeed, taking into account the slow change of the
quadruplet energy with $\varepsilon$, $\tilde{E}_{m,0} \simeq m^2 +
\varepsilon^2/(m^2-1)$, we may directly replace $m$ by
$\sqrt{\tilde{E}_{m,0}}$ in the expression of the splitting. We thus obtain
$\Delta (\tilde{E},\alpha=0) \simeq 4 (\varepsilon/2)^{\sqrt{\tilde{E}}}/\Gamma
(\sqrt{\tilde{E}})^2$ which gives a
rapidly decreasing function of the quadruplet energy for values of
$\varepsilon \ll \tilde{E}$.  

\section{Summary and concluding remarks}

In this paper, after a brief review of the essential results of the single
pendulum problem, we have used a two coupled pendula model to show the
possibility of constructing a quantum state $\ke{\Phi}$ whose properties are
similar to those of a classical rotobreather. This state has been built in a
very similar manner to those used in the corresponding classical 
system, namely, by starting from a state which
mimics the behavior of a rotating pendulum decoupled from another one at
rest at the anti-continuous limit. 
Four states, each belonging to a different symmetry sector, are shown
to be necessary to form it. By switching on the coupling between the
pendula, we follow the resulting evolution of the quadruplet 
and monitor the different splittings. Each of them can be shown to be
associated to a given tunneling process according to the
involved states. 

In general, we have seen that two processes have to be distinguished. The
first concerns 
the tunneling (or transfer) of the initial excitation (energy) between the
pendula which corresponds to a tunneling of the quantum rotobreather
$\ke{\Phi}$ between neighboring sites. This tunneling effect is associated to
the permutation symmetry present in the system.
The second tunneling effect relies on another symmetry due
to the invariance of the system to a reversal of its total momentum. This
symmetry already appears in the single pendulum problem where it is
responsible for the occurrence of doublets of nearly degenerated states at
sufficiently high energy (above the separatrix). This second tunneling effect
takes a time which is  orders
of magnitude larger than the time of the excitation transfer between
pendula (except for very small coupling values). 

By progressively increasing the pendula coupling,
we have shown that the quadruplet under investigation may survive
crossings (or avoided crossings) with other states. Nevertheless,
for large enough coupling, the nearly degenerated states of different
permutation symmetry parities separate sufficiently 
from each other, leading to splittings
of order of the mean level spacing in the spectrum. This situation corresponds
classically to a large chaotization of the phase space at the considered
energy level and thus to a strong overlap of the quantum state with the
chaotic layer surrounding the quasi-regular island where the classical
rotobreather is located. The disappearance of the quantum
rotobreather can be interepreted by the fact 
that the corresponding state is no more able to keep
the excitation in a given pendulum for a long time as compared to the typical 
oscillation time of the system.   

To conclude this paper, we would like to comment on the relation between the
classical and the quantum (roto)breathers. 
An essential ingredient for the existence of a quantum
breather is the appearance of bands of nearly degenerated eigenstates whose
bandwith remain very small as the coupling increases. Nevertheless, the general
method employed to construct the quantum breather by starting from the
anti-continuous limit may be used as well for a chain of $N$ 
purely harmonic oscillators (with a harmonic coupling). 
As this chain cannot possess any kind
of classical breathers, what prevents quantum breathers from existing
on such a chain? To answer this question, we compute
the bandwith corresponding to a local
excitation of $n$ bosons close to the anti-continuous limit. We obtain
$\Delta E_n = n (\sqrt{\omega^2+4\varepsilon}-\omega)$, $\omega$ being the
oscillator frequency and $\varepsilon$ the coupling. This shows that the
bandwith becomes large as the mean energy of the band $E_n \sim n \omega$
does. Moreover, the density of states scales like $g(E) \sim E^{N-1}$ (see the
introduction) and the product $g(E_n) \Delta E_n \rightarrow \infty$ as $E
\rightarrow \infty$. This shows that like the classical harmonic chain, 
the quantum system cannot support quantum breathers.   

A second comment concerns the "classical-like" behavior of the quantum
breather. By defining the latter as a superposition of the
eigenstates making up a single band, its time evolution is de facto restricted
to these states. The fine structure of the band thus provides the only
available frequencies of this evolution. As the corresponding splittings are
very small, none of them is related to the classical breather frequency. 
These frequencies only concern the tunneling of the
excitation from site to site. On the other hand, the individual 
energies of particles making up a classical
breather, for example, oscillate around their mean value at the breather 
frequency. Where does such a frequency appear in the quantum system?
It turns out that this frequency is naturally recovered  
by considering quantum states which display an excitation similar to the
quantum breather but which are not restricted to a single band. 
We have verified in our
system that a coherent state parametrized by the average phases and momenta of
the quantum rotobreather $\ke{\Phi}$ excites mainly the quadruplets separated 
by an energy difference which corresponds approximately
to the classical rotobreather frequency. The interaction between quadruplets
is thus responsible for the "classical-like" time behavior of averaged 
observables (like individual energies). This should generalize to nonlinear 
systems with more than two degrees of freedom where we expect interactions
between bands to play a similar role as the interaction between quadruplets
in our system.

\appendix
\section{Coupled rotors}                 

In this appendix, we solve the problem of two coupled rotors, i.e. the limit
of the two coupled pendula when the onsite parameter $\alpha$ is equal to
zero. In this case, the system defined by the Hamiltonian (\ref{H2pend}) is
integrable. Indeed,
\begin{equation} \label{Hcrxy}
H_{\rm cr} = \frac{1}{2}\left(p_1^2+p_2^2\right)+\varepsilon (1-\cos
(x_1-x_2))
\end{equation}
Let $s=(x_1+x_2)/2$ and $d=(x_1-x_2)/2$. Then the classical equations of
motion reads 
\begin{eqnarray}
&& \ddot{s} = 0 \\
&& \ddot{d} +\varepsilon \sin (2d) = 0 
\end{eqnarray}     
The system decouples and consists of a free rotor for the center of mass motion
and a pendulum for the relative coordinate $2d$ (with an "onsite" parameter
$2\varepsilon$).\\
Let us turn to the associated quantum problem. The Hamiltonian (\ref{Hcrxy})
gives rise to the time independent Schr\"odinger equation
\begin{equation} \label{Schrxy}
\left[ -\frac{1}{2}\left(\partial_{x_1}^2+\partial_{x_2}^2\right)+
\varepsilon (1-\cos (x_1-x_2)) -E\right] \psi(x_1,x_2) = 0
\end{equation}    
which yields (after the change of variables defined above) 
\begin{equation} \label{Schrsd}
\left[ -\frac{1}{4}\left(\partial_{s}^2+\partial_{d}^2\right)+
\varepsilon (1-\cos (2d)) -E\right] \tilde{\psi}(s,d) = 0
\end{equation}
where $\tilde{\psi}(s,d)= \psi(x_1,x_2)$. Because of the $2\pi$-periodicity of
$\psi(x_1,x_2)$ in each of its variables we may expand it as a double Fourier
series:
\begin{eqnarray} \label{Fourxy}
\psi(x_1,x_2) & = &
\frac{1}{2\pi}\sum_{m,n \in \Z^2} \psi_{m,n} \exp(i(mx_1+nx_2))   \\
\label{Foursd} & = &
\frac{1}{2\pi}\left\{ \sum_{\sigma,\delta \in (2\Z)^2}
\tilde{\psi}_{\sigma,\delta}^{(e)} \exp(i(\sigma s+\delta d)) \right.
 \nonumber \\
&+& \left. \sum_{\sigma,\delta \in (2\Z+1)^2}
\tilde{\psi}_{\sigma,\delta}^{(o)} \exp(i(\sigma s+\delta d))
 \right\} \\
& = & \tilde{\psi}(s,d) \nonumber
\end{eqnarray}     
where we have defined $\sigma=m+n$ and $\delta=m-n$ and where the notations
$\sigma,\delta \in (2\Z)^2$ and $\sigma,\delta \in (2\Z+1)^2$ indicate
respectively that $\sigma,\delta$ have to be both 
even numbers or both odd numbers. 
Moreover, $\tilde{\psi}_{\sigma,\delta}^{(e)} =
\psi_{(\sigma+\delta)/2,(\sigma-\delta)/2}$ for $\sigma,\delta$ both even
(reason of the superscript ${}^{(e)}$) and similarly
$\tilde{\psi}_{\sigma,\delta}^{(o)} = 
\psi_{(\sigma+\delta)/2,(\sigma-\delta)/2}$ for $\sigma,\delta$ both odd.
It follows that the first term in (\ref{Foursd}) is a $\pi$-periodic
function in each of the variables $s$ or $d$ whereas the second one is
$2\pi$-periodic. \\
Notice further that because the potential only depends on $d$, (\ref{Schrsd}) 
decouples to give
\begin{eqnarray} \label{decschrsd}
&& \left[ -\frac{1}{2}\partial_{s}^2 - 2E_s\right] \tilde{\psi}_s(s) = 0 \\
&& \left[ -\frac{1}{2}\partial_{d}^2+
2\varepsilon (1-\cos (2d)) - 2E_d \right] \tilde{\psi}_d(d) = 0 
\end{eqnarray} 
where $\tilde{\psi}(s,d) = \tilde{\psi}_s(s) \tilde{\psi}_d(d)$ and
$E=E_s+E_d$. 

The last equation of (\ref{decschrsd}) is a Mathieu equation with
a characteristic value $a$ and a parameter $q$ defined by
\begin{equation}
a = 4(E_d-\varepsilon)\ \ \ \text{and} \ \ \   
q = -2\varepsilon 
\end{equation}
Taking into account the preceeding remarks concerning the periodicity 
as well as the two possible parities (odd or even) of the solutions, we
finally get
\begin{equation}
 E_s^{\sigma} = \frac{1}{4}\sigma^2\ \ \ \text{and}\ \ \ 
E_d^{\delta} = \frac{1}{4}\left(
\begin{array}{c}
   a \\
   b \\
\end{array}
\right)_{\delta}\!\!(q)+\varepsilon
\end{equation} 
where $\sigma,\delta \in \Z^2$ 
have to have the same parity (even or odd) and  
where we have introduced a spinor notation
\begin{equation}\left(
\begin{array}{c}
   f \\
   g \\
\end{array}
\right)_{\mu}\!\!(z) = \left\{ \begin{array}{c}
   f_{\mu}(z) \ \ \ \text{if}\ \ \ \mu \geq 0 \\
   g_{|\mu|}(z) \ \ \ \text{if}\ \ \ \mu < 0  \\
\end{array} \right.
\end{equation}
For the usual notation of the
Mathieu equation see \cite{Abr}. By defining the rescaled energy by
$\tilde{E}=2(E-\varepsilon)$ and using the quantum numbers $m,n$
we finally get the following expression for the spectrum of the coupled rotors
\begin{equation} \label{speccr}
\tilde{E}_{m,n} = \frac{1}{2} \left((m+n)^2+\left(
\begin{array}{c}
   a \\
   b \\
\end{array}
\right)_{(m-n)}\!\!\!\!\!\!\!\!(-2\varepsilon)\right)
\ \ \ \ (m,n) \in \Z^2
\end{equation}
It follows $\tilde{E}_{m,n} = \tilde{E}_{-n,-m}$. Thus all the
levels are twofold degenerated except when $m=-n$, which corresponds to a
total momentum of the system equal to zero. Notice that, when 
$\varepsilon \rightarrow 0$,
the characteristic values converge to $(m-n)^2$ and we recover the free rotors
spectrum $\tilde{E}_{m,n} = m^2+n^2$. Moreover, at sufficiently high
momenta differences $|m-n| \gg \sqrt{\varepsilon}$, the energies
$\tilde{E}_{m,n}$ and $\tilde{E}_{n,m}$ become nearly degenerated. Their
splitting, computed by means of (\ref{ambMat}) gives
\begin{equation}
\Delta E_{|m-n|} = \frac{1}{2} \Delta \tilde{E}_{|m-n|} = 
2 \frac{(\varepsilon/2)^{|m-n|}}{(|m-n|-1)!^2}
\end{equation}
The eigenfunctions corresponding to the spectrum (\ref{speccr}) may be
expressed according to their symmetry sectors by
\begin{equation}
\psi^{(m,n)}(x_1,x_2) = {\cal N}_{\sigma} 
\left(\begin{array}{c} C \\ S \\ \end{array} \right)_{\sigma}\!\!
\left(s\right)
\left(\begin{array}{c} ce \\ se
 \\ \end{array} \right)_{\delta}\!\!
\left(d;q\right)  
\end{equation}
where the normalization factor ${\cal N}_{\sigma}$ is $1/\pi$ for $\sigma
\neq 0$ and $1/\sqrt{2}\pi$ if $\sigma = 0$ and where
 $\sigma$, $\delta$, $s$ and $d$ have already been defined as functions of
$m$, $n$, $x_1$ and $x_2$. $C_{\sigma}(s)=\cos (\sigma s)$ and 
$S_{\sigma}(s)=\sin (\sigma s)$. With this notation, the correspondence
between the symmetry sectors and the values of $\sigma$ and $\delta$ are
\begin{eqnarray}
\ke{s} \Rightarrow (\sigma \geq 0\ ;\ \delta \geq 0)\ \  &;&\ \ 
\ke{a} \Rightarrow (\sigma < 0\ ;\ \delta \geq 0) \nonumber \\
\ke{\bar s} \Rightarrow (\sigma < 0\ ;\ \delta < 0)\ \  &;&\ \ 
\ke{\bar a} \Rightarrow (\sigma \geq  0\ ;\ \delta < 0) \nonumber
\end{eqnarray}

\end{document}